\documentclass[iop,apj]{emulateapj}
\slugcomment{{\sc ApJ, Accepted:} 13 January 2011} 
\usepackage{graphicx}
\usepackage{natbib}
\usepackage{hyperref}
\newcommand{\refeq}[1]{(\ref{#1})}
\newcommand{\qa}{$\phn\phn0^\circ \leq l \leq \phn90^\circ$}
\newcommand{\qb}{$\phn90^\circ \leq l \leq 180^\circ$}
\newcommand{\qc}{$180^\circ \leq l \leq 270^\circ$}
\newcommand{\qd}{$270^\circ \leq l \leq 360^\circ$}
\newlength{\floatwidth}
\begin{document}
\bibliographystyle{apj}
\title{The galaxy counts-in-cells distribution from the SDSS}
\author{Abel Yang}
\affil{Department of Astronomy, University of Virginia, Charlottesville, VA 22904}
\author{William C. Saslaw}
\affil{Institute of Astronomy, Madingley Road, Cambridge CB3 0HA, UK; and Department of Astronomy, University of Virginia, Charlottesville, VA 22904}
\begin{abstract}
We determine the galaxy counts-in-cells distribution from the Sloan Digital Sky Survey~(SDSS) for 3D spherical cells in redshift space as well as for 2D projected cells. We find that cosmic variance in the SDSS causes the counts-in-cells distributions in different quadrants to differ from each other by up to $20\%$. We also find that within this cosmic variance, the overall galaxy counts-in-cells? distribution agrees with both the gravitational quasi-equilibrium distribution and the negative binomial distribution. We also find that brighter galaxies are more strongly clustered than if they were randomly selected from a larger complete sample that includes galaxies of all luminosities. The results suggest that bright galaxies could be in dark matter haloes separated by less than $\sim10 h^{-1}$ Mpc.
\end{abstract}

\keywords{galaxies: statistics --- cosmology: theory --- large-scale structure
of universe --- gravitation}

\section{Introduction}
\label{sec-intro}
The galaxy counts-in-cells distribution is a simple but powerful statistic which characterizes the locations of galaxies in space. It includes statistical information on voids and other underdense regions, on clusters of all shapes and sizes, on filaments, on the probability of finding an arbitrary number of neighbors around randomly located positions, on counts of galaxies in cells of arbitrary shapes and sizes randomly located, and on galaxy correlation functions of all orders. These are just some of its representations~\citep{2000dggc.book.....S, 2009arXiv0902.0747S}. Moreover it is also closely related to the distribution function of the peculiar velocities of galaxies around the Hubble flow~\citep{1990ApJ...365..419S, 2004ApJ...608..636L}.

Although the counts-in-cells distribution contains a large amount of information about galaxy clustering, it has not received as much attention as more common statistical descriptions of clustering such as the two-point correlation function. In addition, most earlier studies have focused on the counts-in-cells distribution for a magnitude-limited sample in projection~(e.g. \citealt{2005ApJ...626..795S} and references within). While there have been studies that have used redshift-limited samples~(e.g. \citealt{1998ApJ...509..595S, 2009ApJ...695.1121R}), their samples were generally smaller and their statistics were less precise.

Other studies have also examined the void probability function, which is a special case of the counts-in-cells distribution that describes the distribution of the volumes of voids, or regions with no galaxies. However, studies~\citep{1984ApJ...276...13S,1986ApJ...306..358F} have shown that the void probability function can be entirely described by the volume integral of the two-point correlation function $\overline{\xi}_2$ and the mean number of galaxies in a cell $\overline{N}$. This suggests that the void probability function alone is insufficient to completely describe the clustering of galaxies. To do so we would have to consider more than just voids.

Various statistical descriptions for the distribution function have been developed~(for an early review see \citealt{1986ApJ...306..358F}) with the gravitational quasi-equilibrum distribution~(GQED, \citealt{1984ApJ...276...13S, 2002ApJ...571..576A}) and the negative binomial distribution~(NBD, \citealt{1992MNRAS.254..247E, 1995MNRAS.274..213S}) in common use. While the GQED can be derived from thermodynamics~\citep{1984ApJ...276...13S} and statistical mechanics~\citep{2002ApJ...571..576A}, the NBD has been shown to violate the second law of thermodynamics by \citet{1996ApJ...460...16S}.

Observations however show a more complex picture. While the counts-in-cells distribution for the 2MASS catalog in projection shows a good agreement with the GQED~\citep{2005ApJ...626..795S}, an analysis of the void probability function for the SDSS and DEEP2 catalogs in redshift space by \citet{2005ApJ...635..990C} suggests a closer agreement with the NBD. This disagreement between projection and redshift space complicates our understanding of the theory behind the clustering of galaxies and raises a number of questions. Why does the observed counts-in-cells distribution agree with the NBD in some cases and the GQED in others? What are the conditions under which the counts-in-cells distribution agrees more closely with the GQED or NBD? Moreover, should the universe be allowed to violate the second law of thermodynamics?

In section 2 we describe the distribution functions and the information they contain. In particular, we describe the derivation and some aspects of the GQED and NBD. In section 3, we describe the procedure used to measure the counts in cells distribution from the SDSS NYU-VAGC catalog~\citep{2005AJ....129.2562B}. In section 4 we present our results for the 2-point correlation function and $f_V(N)$. In section 5 we summarize our findings. Following \citet{2005AJ....129.2562B}, we use $\Omega_m = 0.3$, $\Omega_k = 0.0$, $\Omega_\Lambda = 0.7$ and $H_0 = 100 h$ km s$^{-1}$ Mpc$^{-1}$.

\section{Distribution functions}
The most general form of the counts-in-cells distribution is denoted by $f(N,V)$ which gives the probability of finding $N$ galaxies in a region of space with volume $V$. There are two approaches to studying this distribution. The first approach is to let $V$ be constant resulting in $f_V(N)$ which gives the distribution of the number of galaxies $N$ for cells of a given volume $V$. This method is simple to use, yet powerful. The measurement of $f_V(N)$ generally involves examining cells in 3D space or in projection and counting the number of galaxies in each cell.

In addition, the moments of $f_V(N)$ are closely related to the volume integrals of the correlation functions of all orders (e.g. \citealt{1980lssu.book.....P, 1985ApJ...289...10F, 2000dggc.book.....S}) and the correlation functions can be measured from the moments of $f_V(N)$~\citep{1994ApJ...425....1F}. For example the relation between the volume integrals of the 2-point and 3-point correlation functions and the moments of the counts-in-cells distribution are given by
\begin{eqnarray}\label{edist-corr}
\langle(\Delta N)^2\rangle &=& \overline{N}^2\overline{\xi}_2 + \overline{N} \\
\langle(\Delta N)^3\rangle &=& \overline{N}^3\overline{\xi}_3 + 3\overline{N}^2\overline{\xi}_2 + \overline{N}
\end{eqnarray}
where $\overline{N}$ is the mean number of galaxies in a cell and the volume integral of the $N$-point correlation function
\begin{equation}\label{edist-vcorr}
\overline{\xi}_N(V) = \frac{1}{V^N} \int_V \xi_N(\mathbf{r}_1, \ldots, \mathbf{r}_N)d^3\mathbf{r}_1\ldots d^3\mathbf{r}_N
\end{equation}
with $\overline{\xi}_1 = 1$ depends on the cell volume $V$. This property allows us to compare the counts-in-cells results with observations of the two-point correlation function.

To get the measured value of the two-point correlation function $\xi_2(r)$ we rewrite equation \refeq{edist-vcorr} for the 2-galaxy case as
\begin{equation}\label{edist-2ptcorr}
\overline{\xi}_2(r) = \frac{1}{V(r)} \int_0^r \frac{dV}{dr} \xi_2(r) dr.
\end{equation}
This is a conditional average correlation where one galaxy is located at the center of the volume so one power of $V$ in the denominator is removed by using polar coordinates relative to the central galaxy of the arbitrary volume.

We can invert the integral using a finite difference scheme with an interval of $\Delta r$ to approximate the value of $\xi_2(r)$ such that
\begin{equation}\label{edist-2ptvar}
\xi_2(r) = \frac{\overline{\xi}_2(r+\Delta r)V(r+\Delta r) - \overline{\xi}_2(r)V(r)}{V(r+\Delta r) - V(r)}
\end{equation}
where from equation \refeq{edist-corr}
\begin{equation}\label{edist-xibvar}
\overline{\xi}_2 =\frac{\langle(\Delta N)^2\rangle-\overline{N}}{\overline{N}^2}
\end{equation}
and $V$ is the volume of the cell which depends on the shape of the cell. This gives us a means of determining the two-point correlation function from a series of measurements of $f_V(N)$ over a range of scales.

The other approach to studying $f(V,N)$ is to let $N$ be constant resulting in $f_N(V)$ which gives the distribution of the volume $V$ occupied by $N$ galaxies of which the void probability function~(VPF), where $N=0$, is a special case~(e.g. \citealt{1986ApJ...301....1C}). A theoretical approach to $f_N(V)$ is complicated by the fact that the distribution depends on the correlation function at all scales rather than a scale determined by a particular value of $V$. This scale dependence can be found either empirically from the dependence of the variance of the $f_V(N)$ distribution on $V$, or from a model assumption of the form of $\overline{\xi}_2(V)$. These give the analytic form of $f_N(V)$. To avoid these complications, most attempts to study $f_N(V)$ have focused on the VPF because use of the reduced void probability~\citep{1986ApJ...306..358F} considerably simplifies the analysis by expressing $f_0(V)$ in terms of $\overline{N}\overline{\xi}_2$.

The reduced void probability, given by
\begin{equation}\label{edist-rvp}
\chi(\overline{N}\overline{\xi}_2) \equiv -\frac{\ln(f_0(V))}{\overline{N}},
\end{equation}
provides a means of isolating the scale-dependence of the void probability function because $\chi$ is a function that depends only on $\overline{N}\overline{\xi}_2$, and $\overline{N}\overline{\xi}_2$ is easily derived from the variance of $f_V(N)$. However, this simplification is only possible for voids in the GQED and NBD because, for $N \geq 1$, $f_N(V)$ depends on $\overline{N}$ and $\overline{\xi}_2$ separately. Moreover, the void distribution is relatively insensitive to information on large cell sizes because large cells are unlikely to be completely empty. For these reasons we focus on the simpler $f_V(N)$ approach in this paper and introduce the statistical descriptions of the counts-in-cells distribution.

\subsection{The GQED}
The gravitational quasi-equilibrium distribution was first derived from thermodynamics~\citep{1984ApJ...276...13S} and subsequently from statistical mechanics~\citep{2002ApJ...571..576A} by assuming that galaxy clustering evolves through a sequence of quasi-equilibrium states. The resulting distribution is given by
\begin{equation}\label{edist-GQED}
f_{V,GQED}(N) = \frac{\overline{N}(1-b)}{N!}
\left(\overline{N}(1-b)+Nb\right)^{N-1} e^{-\overline{N} (1-b)-Nb}
\end{equation}
where $\overline{N}=\overline{n}V$ is the average expected number of galaxies in a cell of volume $V$ and $\overline{n}$ is the average number density of galaxies. Here $b=-W/2K$ is the ratio of the gravitational correlation energy $W$ to twice the kinetic energy $K$ of peculiar velocities relative to the Hubble flow and it represents a measure of clustering.

A physical description of $b$ is given by \citet{2002ApJ...571..576A} to be
\begin{equation}\label{edist-bdef}
b = \frac{3/2(Gm^2)^3 \overline{n}T^{-3}}{1+3/2(Gm^2)^3 \overline{n}T^{-3}}
\end{equation}
which relates $b$ to the mass of a galaxy $m$, the number density of galaxies $\overline{n}$ and the kinetic temperature of the galaxies $T$. Here $G$ is the gravitational constant. Originally an ansatz proposed by \citet{1984ApJ...276...13S}, the physical origin of $b$ was only later understood through work done by \citet{1996ApJ...460...16S} on the first and second laws of thermodynamics, and through the statistical mechanical derivation of the GQED by \citet{2002ApJ...571..576A}.

We can relate the clustering parameter $b$ to the variance of the counts-in-cells distribution through
\begin{equation}\label{edist-bvar}
\langle(\Delta N)^2\rangle = \frac{\overline{N}}{(1-b)^2}.
\end{equation}
which allows us to describe the clustering of galaxies with the GQED in a self-consistent manner with no free parameters. This also allows us to relate $b$ to the volume integral of the two-point correlation function such that
\begin{equation}\label{edist-bxi}
b = 1 - \left(\overline{N}\overline{\xi}_2(V) + 1\right)^{-1/2}
\end{equation}
which indicates that $b$ depends on $\overline{\xi}_2$ and varies with cell volume $V$.

Although the derivation of equation \refeq{edist-GQED} by \citet{2002ApJ...571..576A} was done assuming that all galaxies have the same mass, theoretical work by \citet{2006IJMPD..15.1267A} showed that the statistical mechanical framework can be extended to take into account population components of differing masses. In addition, $N$-body simulations by \citet{1993ApJ...403..476I} also showed that the GQED for the case where galaxies are of the same mass is often a good fit to $N$-body simulations where galaxies are allowed to take a range of masses. This suggests that the GQED given in equation \refeq{edist-GQED} is a reasonable approximation to the counts-in-cells distribution. Together with the physical motivation behind its derivation, the GQED can be used to gain further insights into the physics behind the counts-in-cells distribution.

\subsection{The NBD}
The negative binomial distribution was proposed in the cosmological context by \citet{1983PhLB..131..116C} and subsequently derived by \citet{1992MNRAS.254..247E} by describing the distribution as a statistical random process where $N$ galaxies are introduced in $m$ spatially disconnected boxes. In this model, the probability that a galaxy is introduced in a particular box is proportional to the number of galaxies already inside the box. The resulting distribution is 
\begin{equation}\label{edist-NBD}
f_{V,NBD}(N) =
\frac{\Gamma\left(N+\frac{1}{g}\right)}{\Gamma\left(\frac{1}{g}\right)N!} \frac{\overline{N}^N \left(\frac{1}{g}\right)^\frac{1}{g}} {\left(\overline{N}+\frac{1}{g}\right)^{N+\frac{1}{g}}}
\end{equation}
where
\begin{equation}\label{edist-gxi}
g = \overline{\xi}_2(V)
 = \frac{\langle(\Delta N)^2\rangle - \overline{N}}{\overline{N}^2}
\end{equation}
is a clustering parameter that depends on cell volume and $\Gamma$ is the standard gamma function. Similar to the GQED, the NBD can also describe the counts-in-cells distribution self-consistently with no free parameters, and the clustering parameter $g$ is just $\overline{\xi}_2$. 

An alternative derivation of the NBD in the thermodynamic framework of \citet{1984ApJ...276...13S} is given by \citet{1995MNRAS.274..213S}. In this case, the equivalent of $b$ is given by
\begin{equation}\label{edist-nbbdef}
b = 1-\frac{\ln(1+b_0\overline{n}T^3)}{b_0\overline{n}T^3}.
\end{equation}
Although this form fulfils $0 \leq b \leq 1$, it was found to violate the second law of thermodynamics by \citet{1996ApJ...460...16S} which suggests that the NBD is not physically motivated. A closer look at the statistical random process from which the NBD was derived suggests that the NBD assumes galaxies form where there is already a cluster of galaxies. This process does not take infall into account, and hence the depletion of regions outside a cluster that occur in the process of infall is not taken into account.

From the derivation of the NBD by \citet{1992MNRAS.254..247E}, we note that the NBD can describe the case where galaxies form from the merger of less massive objects. In this description, the less massive objects can be expected to follow the GQED, but not all of them can be observed. These objects may merge to form objects bright enough to be observed, and their locations are likely to be in denser regions that contain a higher density of fainter objects. $N$-body simulations by \citet{2005ApJ...635..990C} show that while the VPF for galaxies follows the NBD, the VPF for dark matter particles follows the GQED. While this qualitative explanation may seem plausible, a detailed quantitative analysis will depend on the physics of the more complicated halo occupation distribution.

\section{Data and Procedure}

\subsection{Catalog data}
The New York University value-added galaxy catalog~(NYU-VAGC, \citealt{2005AJ....129.2562B}) is a composite catalog with the Sloan Digital Sky Survey~(SDSS) data as its primary component. It contains over 550,000 galaxies with their redshifts and positions on the sky. The catalog also contains extinction corrected and $K$-corrected absolute magnitudes for 8 bands, of which the $u$, $g$, $r$, $i$ and $z$ bands come from the SDSS and the $J$, $H$ and $K_s$ bands come from the 2-Micron All-Sky Survey~(2MASS) although for this study we use only the data from the SDSS. The galaxies in the catalog are also corrected for fiber collisions using the ``nearest'' method described in \citet{2005AJ....129.2562B}. Less that $10\%$ of the galaxies are affected by this correction which allows for a more complete sample in crowded regions.

In addition to the galaxy catalog, the NYU-VAGC also contains a survey geometry catalog that describes the survey footprint in terms of spherical polygons~(described in \citealt{2005AJ....129.2562B}). Since the SDSS is not an all-sky survey, the survey footprint determines the positions of cells and allows us to lay down cells where there is valid data.

For this work, we use the large scale structure samples in the version of the catalog corresponding to the seventh data release of the SDSS~\citep[DR7]{2009ApJS..182..543A}. We use the subsample with a flux limit of $r < 17.6$ and perform further selection cuts based on the properties of the sample. In particular, we choose absolute magnitude cuts to obtain a complete sample within a given redshift range.

We consider two redshift ranges in the $g$, $r$ and $i$ bands at $0.04 \leq z \leq 0.12$ and $0.12 \leq z \leq 0.20$. The low redshift limit of $z \geq 0.04$ ensures that the sample is within the Hubble flow, and excludes the Coma and Virgo clusters. Since the SDSS ``great wall'' spans a redshift range of $0.065 \leq z \leq 0.09$~\citep{2005ApJ...624..463G}, it is fully contained within the low redshift range. This allows us to isolate the effect of the ``great wall'' by comparing the low redshift range to the high redshift range.

To determine a suitable absolute magnitude cut, we define the faint limit $M_f$ as the absolute magnitude where the observed luminosity function begins to turn over because the limiting magnitude has been reached. This means that for a faint limit $M_f$ and limiting redshift $z_{max}$, the comoving number density of galaxies $\overline{n}(M_f)$ brighter than $M_f$ should be the approximately the same for any limiting redshift $z < z_{max}$.

We obtain $M_f$ by comparing $\overline{n}(M_f)$ for a redshift range with a lower redshift subset of the same range. For the low range, we compare the range $0.04\leq z\leq 0.12$ with the range $0.04\leq z\leq 0.10$ and for the high redshift range we compare the range $0.12\leq z\leq 0.20$ with the range $0.12\leq z\leq 0.18$. The optimal faint limit $M_f$ which gives us the largest complete sample occurs where the compared values of $\overline{n}(M_f)$ are approximately equal.

\begin{figure}
\begin{center}
\includegraphics[width=\floatwidth]{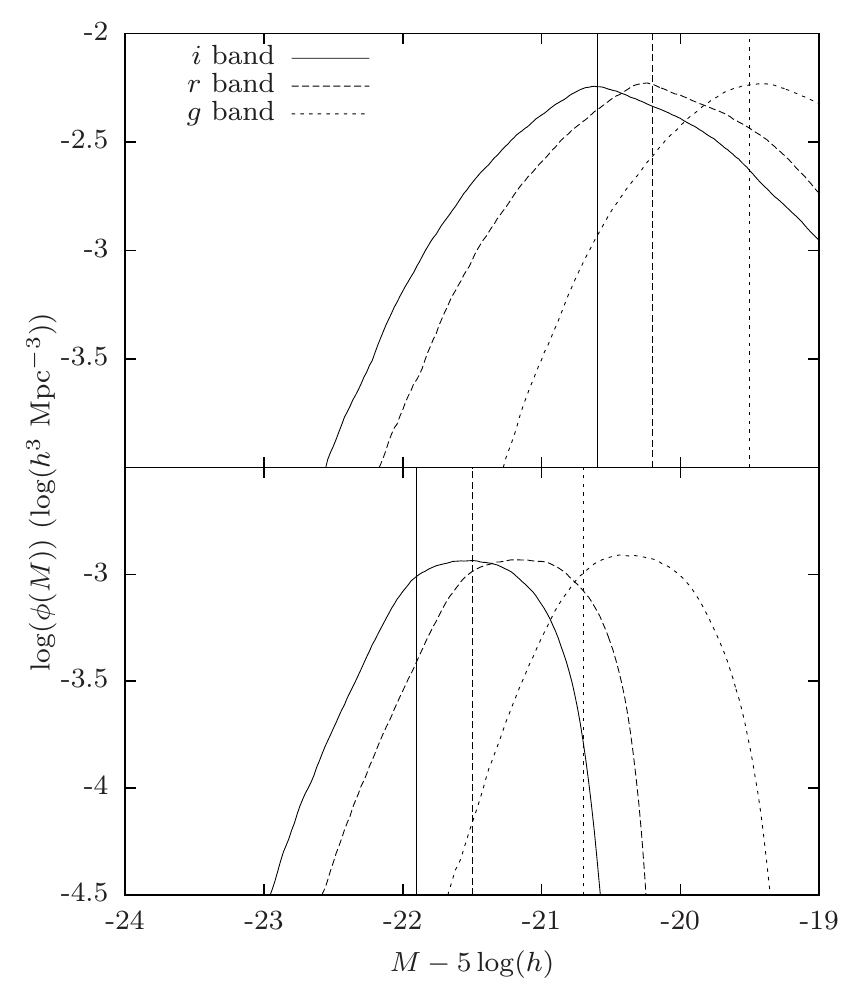}
\end{center}
\caption{\label{fdata-lf}
Observed luminosity function for the NYU-VAGC at $0.04 \leq z \leq 0.12$~(top panel) and $0.12 \leq z \leq 0.20$~(bottom panel). The vertical lines indicate the absolute magnitude cuts we have adopted.}
\end{figure}

We find that the lower redshift range is complete for $M_g < -19.5$, $M_r < -20.2$ and $M_i < -20.6$ while the higher redshift range is complete for $M_g < -20.7$, $M_r < -21.5$ and $M_i < -21.9$. We plot these limits on the observed luminosity function at $0.04 \leq z \leq 0.12$ and $0.12 \leq z \leq 0.20$ in figure \ref{fdata-lf} and summarize the subsamples we use in table \ref{tdata-zcut}.

\begin{deluxetable}{l c c c c}
\tablewidth{0pt}
\tablewidth{\floatwidth}
\tablecaption{\label{tdata-zcut}
Selected subsamples
}
\tablehead{
\colhead{Sample} & \colhead{Redshift} & \colhead{Magnitude} & \colhead{Density $\overline{n}$} & \colhead{Galaxies} \\
 & & \colhead{$M-5\log(h)$} &\colhead{$h^{-3}$ Mpc$^3$} &
}
\startdata
1a(g) & $0.04$ -- $0.12$ & $M_g < -19.5$ & $4.22\times 10^{-3}$ & $132977$ \\
1a(r) & $0.04$ -- $0.12$ & $M_r < -20.2$ & $4.65\times 10^{-3}$ & $146413$ \\
1a(i) & $0.04$ -- $0.12$ & $M_i < -20.6$ & $4.41\times 10^{-3}$ & $138754$ \\
\\
1b(g) & $0.04$ -- $0.12$ & $M_g < -20.7$ & $2.38\times 10^{-4}$ & $7483$ \\
1b(r) & $0.04$ -- $0.12$ & $M_r < -21.5$ & $2.78\times 10^{-4}$ & $8737$ \\
1b(i) & $0.04$ -- $0.12$ & $M_i < -21.9$ & $2.63\times 10^{-4}$ & $8263$ \\
\\
2b(g) & $0.12$ -- $0.20$ & $M_g < -20.7$ & $3.61\times 10^{-4}$ & $39772$ \\
2b(r) & $0.12$ -- $0.20$ & $M_r < -21.5$ & $3.97\times 10^{-4}$ & $43725$ \\
2b(i) & $0.12$ -- $0.20$ & $M_i < -21.9$ & $3.68\times 10^{-4}$ & $40521$ \\
\enddata
\end{deluxetable}

Here we note that the 1a(g), 1a(r) and 1a(i) samples have similar spatial densities, and likewise the 1b(g), 1b(r) and 1b(i), and 2b(g), 2b(r) and 2b(i) samples also have similar spatial densities. Hence any large differences in clustering between the samples of different color should arise from selection effects that depend on color.

\subsection{Cosmic Variance}

An analysis by \citet{2009A&A...508...17S} found systematic variations between different subvolumes of the SDSS catalog on scales larger than $30 h^{-1}$ Mpc such that these subvolumes are not statistically similar. These variations are likely to be caused by cosmic variance that our analysis should take into account. We use two approaches to analyze the effect of cosmic variance on our results.

The first is simply to consider independent subfields of the survey footprint. To ensure that effects of galactic latitude, distance and lookback time are constant across all subsamples, we compute and compare the counts-in-cells distribution for cells in non-overlapping quadrants in galactic longitude. The size of quadrants is also comparable with the size of the SDSS ``great wall'' which spans $\sim70^\circ$~\citep{2006ChJAA...6...35D}. In this approach, all cells that belong to a quadrant are fully-contained within the selected quadrant. This gives us a picture of the variations among widely separated areas of the sky. We consider subsamples in four quadrants such that for galactic longitude $l$, quadrant 1~(q1) covers $0.0^\circ \leq l \leq 90.0^\circ$, quadrant 2~(q2) covers $90.0^\circ \leq l \leq 180.0^\circ$, quadrant 3~(q3) covers $180.0^\circ \leq l \leq 270.0^\circ$ and quadrant 4~(q4) covers $270.0^\circ \leq l \leq 360.0^\circ$. For some samples, we also examine the quadrant to quadrant variations where the quadrant boundaries boundaries have been shifted by $30^\circ$, $45^\circ$ and $60^\circ$ in galactic longitude to check that any variation between quadrants is not caused by our choice of quadrant boundaries.

Moreover, we have varied the size of subsamples and found that for smaller subsamples, \textit{e.g.} sixths rather than quadrants, the subsample to subsample variations are smaller than for larger subsamples. However, subsamples that are too large will not be independent, and there will be too few of them to provide an accurate estimate of the cosmic variance between disconnected regions of the sky. Therefore quadrants are a reasonable subsample size to use.

The second approach is a jackknife-style approach where we leave out cells that fall within a region of the sky, selected based on a quasi-random sequence from \citet{1994ACMTMS-20-494}\footnote{Implemented in the GNU Scientific Library (\url{http://www.gnu.org/software/gsl/})} such that each part of the survey region is equally likely to be chosen for exclusion. For our analysis, we use 1000 different exclusion regions which are circular with a radius of $15^\circ$. This corresponds to a transverse distance of about $60 h^{-1}$ Mpc at $z \sim 0.04$ and an area of approximately 10\% of the SDSS footprint for a ``leave 10\% out'' jackknife procedure from which we can determine the 1-$\sigma$ error.

Here we wish to stress that the jackknife errors are only valid in the case where the SDSS catalog is a representative sample of the universe. This condition essentially requires the universe to be statistically homogenous at scales larger than the SDSS footprint. If this requirement is not met, the errors will not be meaningful because the sample is not a representative sample of the universe.

\subsection{Counts-in-cells strategy}

To obtain the counts-in-cells distribution, we take a sample of cells whose positions are evenly distributed over the survey footprint. To sample the galaxies efficiently, we use a number of cells approximately equal to the number of galaxies, so based on table \ref{tdata-zcut} we use a sample of approximately $130,000\sim140,000$ cells.

To ensure that the entire survey footprint is sampled without bias to any particular region of space, we use the following procedure. We first define an ``instance'' as a set of cells that tile the entire survey region with a consistent amount of overlap. For small cell sizes, a single instance provides enough cells for reliable statistics. For larger cell sizes, we use multiple instances for efficient sampling. To do so, we displace the origin of each subsequent instance by a quasi-random sequence from \citet{1994ACMTMS-20-494}\footnotemark[1] so that cells from no two instances exactly line up with each other. Because we may be dealing with cells on small scales and because cells are allowed to overlap, adjacent cells are generally not independent and hence we cannot use statistical tests which assume samples are independent.

Since the SDSS is not an all-sky survey, we also check each cell against the survey coverage area. We express the projected extent of each cell in terms of spherical polygons and check these against the survey geometry of the NYU-VAGC. We use the \textsc{mangle} software package~\citep{2004MNRAS.349..115H,2008MNRAS.387.1391S} to combine the survey geometry and the bright star mask into a combined exclusion mask that removes areas that are either not in the SDSS or are obscured by a foreground object. In addition, we also discard regions with a galactic latitude below $20^\circ$ to further minimize foreground contamination from the galaxy.

To accept or reject a cell, we use \textsc{mangle} to compute the overlap between the combined exclusion mask and the cell. Since cells may be partly obscured by foreground objects even if they are well within the contiguous region of the survey, we accept cells that have less than $5\%$ of their area masked out. This allows cells to have small regions that may be obscured by foreground objects while having a minor effect of less than 5\% on our statistics.

The counts-in-cells distribution $f_V(N)$ is then obtained by taking the histogram of the number of galaxies within a cell. For this study, we consider 2-dimensional circular cells projected on the sky and 3-dimensional spherical cells in redshift space.

\subsubsection{2-dimensional cells projected on the sky}
For the study of the 2-dimensional projected cells, we use circular cells because their areas and membership are simple to calculate. Such cells can be represented by only two parameters, the cell radius $\theta$ and the position of the cell center. This allows the cell to be described as a spherical polygon with only one cap which is easily processed by \textsc{mangle}. The area of a circular cell in steradians is given by $2\pi(1-\cos \theta)$, and a galaxy is a member of a cell if the great circle distance between the galaxy and the cell center is less than $\theta$.

With a redshift limited sample defined such that the redshift $z$ falls in the range $z_1 \leq z \leq z_2$, we can also determine the comoving volume of the cell with angular radius $\theta$ using
\begin{equation} \label{edata-2dcvol}
V(\theta;z_1, z_2) = \frac{2(1-\cos \theta)}{3} \left[ D(z_2)^3 - D(z_1)^3\right] 
\end{equation}
where the comoving distance $D(z)$ is given by integrating the Friedmann equation
\begin{equation} \label{edata-zdist}
D(z) = \int_{0}^{z} \frac{cz'}{H_0}\left(\Omega_m(1+z')^3 + \Omega_k(1+z')^2 + \Omega_\Lambda \right)^{-1/2}dz'.
\end{equation}

To obtain the 2D projected counts-in-cells distribution, we first map the celestial sphere onto an equal-area sinusoidal projection using
\begin{eqnarray} \label{edata-sproj}
x_0 &=& \alpha \cos(\delta) \nonumber \\
x_1 &=& \delta
\end{eqnarray}
where $\alpha$ and $\delta$ refer to the J2000.0 right ascension and declination respectively. For each instance, we place cell centers on a square grid overlaid on this projection at intervals of $\sqrt{2} \theta$. Subsequent projections will have $x_0$ and $x_1$ shuffled by an amount less than $\sqrt{2} \theta$. For the 2D sample we consider cells with radii between $0.05^\circ$ and $6.0^\circ$ in steps of $0.05^\circ$.

\subsubsection{3-dimensional cells in redshift space}
For 3-dimensional cells in redshift space, we use spherical cells because they are simple to analyse. For example, the simplest form of equation \refeq{edist-vcorr} applies to spherical cells, and such cells can be described by just their location and radius $r$.

To obtain the 3D counts-in-cells distribution in redshift space, we first convert redshift space into Cartesian coordinates using (c.f. \citealt{2005AJ....129.2562B})
\begin{eqnarray} \label{edata-xcart}
x_0 &=& D \cos \delta \cos \alpha \nonumber \\
x_1 &=& D \cos \delta \sin \alpha \nonumber \\
x_2 &=& D \sin \delta
\end{eqnarray}
where $\alpha$ and $\delta$ refer to the J2000.0 right ascension and declination respectively, and $D$ is the comoving distance given by equation \refeq{edata-zdist}. For each instance, we place cell centers on a cartesian grid with spacings of $\sqrt{2} r$. Subsequent instances will have the origin of the grid shuffled by an amount less than $\sqrt{2} r$. For the 3D sample we consider cells with radii between $2.0 h^{-1}$ Mpc and $36.0 h^{-1}$ Mpc in steps of $0.2 h^{-1}$ Mpc.

Since we work in comoving coordinates, the resulting projected area of a cell is a circle about the cell center of angular radius $\theta = \sin^{-1}(r/D)$ where $r$ is the radius of the cell in comoving coordinates. The cell center is also easily obtained from $x_0$, $x_1$ and $x_2$. Hence, we can define a spherical polygon that represents the footprint of the cell on the sky in a manner similar to what we have used for the case of 2D cells.

To get the positions of galaxies in redshift space, we apply the transformation in equation \refeq{edata-xcart}. Then a galaxy is a member of a cell if the distance between the galaxy and cell center is less than $r$.

\section{Results}

\subsection{The Two-Point Correlation Function}
Since the two-point correlation function is a well-studied description of clustering, we first compute the two-point correlation functions $\xi_2(r)$ from the counts-in-cells distribution using equation \refeq{edist-xibvar} and compare our results with earlier works. This allows us to check the validity of our data and method by comparing our results to results from previous studies.

For this study, we focus on the power law approximation of the two-point correlation function since we are dealing with small scales. The power law approximation of the two-point correlation function at small scales is~(e.g. \citealt{1969PASJ...21..221T})
\begin{equation} \label{eres-3dcorr}
\xi_{2,3D}(r) = \left(\frac{r}{r_0}\right)^{-\gamma}
\end{equation}
for 3D cells and
\begin{equation} \label{eres-2dcorr}
\xi_{2,2D}(\theta) = \left(\frac{\theta}{\theta_0}\right)^{-\gamma+1}
\end{equation}
for 2D cells. We obtain the parameters $r_0$ and $\gamma$ by fitting a linear relation between $\log(\xi_2(r))$ and $\log(r)$.

\begin{deluxetable}{l c cc}
\tablewidth{\floatwidth}
\tablewidth{0pt}
\tablecaption{\label{tres-2dcorr}
Two-point correlation function $\xi_{2,2D}(\theta)$ for 2D cells
}
\tablehead{
\colhead{Sample} & \colhead{Quadrant}
 & \colhead{$\theta_0$($^\circ$)} & \colhead{$\gamma$} 
}
\startdata
1a(g) & All & $0.053_{-0.001}^{+0.002}$ & $1.68_{-0.02}^{+0.02}$ \\
1a(g) & \qa & $0.046_{-0.004}^{+0.006}$ & $1.56_{-0.02}^{+0.04}$ \\
1a(g) & \qb & $0.064_{-0.003}^{+0.007}$ & $1.78_{-0.05}^{+0.09}$ \\
1a(g) & \qc & $0.057_{-0.004}^{+0.007}$ & $1.81_{-0.05}^{+0.08}$ \\
1a(g) & \qd & $0.056_{-0.006}^{+0.008}$ & $1.79_{-0.11}^{+0.15}$ \\
\\
1a(r) & All & $0.066_{-0.001}^{+0.002}$ & $1.71_{-0.02}^{+0.02}$ \\
1a(r) & \qa & $0.063_{-0.006}^{+0.007}$ & $1.59_{-0.02}^{+0.04}$ \\
1a(r) & \qb & $0.075_{-0.003}^{+0.009}$ & $1.80_{-0.05}^{+0.09}$ \\
1a(r) & \qc & $0.065_{-0.004}^{+0.007}$ & $1.83_{-0.05}^{+0.08}$ \\
1a(r) & \qd & $0.069_{-0.007}^{+0.007}$ & $1.82_{-0.11}^{+0.15}$ \\
\\
1a(i) & All & $0.068_{-0.001}^{+0.002}$ & $1.72_{-0.02}^{+0.02}$ \\
1a(i) & \qa & $0.066_{-0.007}^{+0.007}$ & $1.60_{-0.02}^{+0.04}$ \\
1a(i) & \qb & $0.078_{-0.002}^{+0.009}$ & $1.81_{-0.04}^{+0.09}$ \\
1a(i) & \qc & $0.068_{-0.004}^{+0.006}$ & $1.84_{-0.05}^{+0.08}$ \\
1a(i) & \qd & $0.074_{-0.006}^{+0.007}$ & $1.83_{-0.11}^{+0.15}$ \\
\\
\\
1b(g) & All & $0.140_{-0.002}^{+0.005}$ & $1.74_{-0.02}^{+0.02}$ \\
1b(g) & \qa & $0.159_{-0.015}^{+0.025}$ & $1.62_{-0.02}^{+0.07}$ \\
1b(g) & \qb & $0.174_{-0.005}^{+0.034}$ & $1.95_{-0.06}^{+0.11}$ \\
1b(g) & \qc & $0.113_{-0.016}^{+0.005}$ & $1.81_{-0.06}^{+0.04}$ \\
1b(g) & \qd & $0.113_{-0.015}^{+0.023}$ & $1.75_{-0.13}^{+0.17}$ \\
\\
1b(r) & All & $0.169_{-0.003}^{+0.005}$ & $1.76_{-0.03}^{+0.02}$ \\
1b(r) & \qa & $0.172_{-0.020}^{+0.032}$ & $1.61_{-0.02}^{+0.04}$ \\
1b(r) & \qb & $0.212_{-0.005}^{+0.030}$ & $2.07_{-0.05}^{+0.08}$ \\
1b(r) & \qc & $0.141_{-0.019}^{+0.007}$ & $1.87_{-0.08}^{+0.04}$ \\
1b(r) & \qd & $0.148_{-0.013}^{+0.026}$ & $1.73_{-0.14}^{+0.18}$ \\
\\
1b(i) & All & $0.168_{-0.003}^{+0.005}$ & $1.78_{-0.02}^{+0.02}$ \\
1b(i) & \qa & $0.173_{-0.015}^{+0.026}$ & $1.66_{-0.02}^{+0.04}$ \\
1b(i) & \qb & $0.200_{-0.007}^{+0.036}$ & $1.98_{-0.05}^{+0.09}$ \\
1b(i) & \qc & $0.147_{-0.018}^{+0.008}$ & $1.90_{-0.08}^{+0.05}$ \\
1b(i) & \qd & $0.146_{-0.012}^{+0.020}$ & $1.73_{-0.14}^{+0.17}$ \\
\\
\\
2b(g) & All & $0.051_{-0.002}^{+0.004}$ & $1.85_{-0.02}^{+0.05}$ \\
2b(g) & \qa & $0.052_{-0.008}^{+0.004}$ & $1.92_{-0.07}^{+0.07}$ \\
2b(g) & \qb & $0.051_{-0.006}^{+0.012}$ & $1.92_{-0.07}^{+0.14}$ \\
2b(g) & \qc & $0.044_{-0.005}^{+0.016}$ & $1.70_{-0.05}^{+0.24}$ \\
2b(g) & \qd & $0.070_{-0.011}^{+0.017}$ & $2.05_{-0.09}^{+0.15}$ \\
\\
2b(r) & All & $0.063_{-0.002}^{+0.004}$ & $1.86_{-0.03}^{+0.05}$ \\
2b(r) & \qa & $0.060_{-0.006}^{+0.004}$ & $1.95_{-0.06}^{+0.06}$ \\
2b(r) & \qb & $0.063_{-0.008}^{+0.019}$ & $1.94_{-0.09}^{+0.20}$ \\
2b(r) & \qc & $0.057_{-0.010}^{+0.016}$ & $1.72_{-0.06}^{+0.19}$ \\
2b(r) & \qd & $0.080_{-0.009}^{+0.019}$ & $2.02_{-0.03}^{+0.11}$ \\
\\
2b(i) & All & $0.066_{-0.002}^{+0.004}$ & $1.87_{-0.03}^{+0.05}$ \\
2b(i) & \qa & $0.063_{-0.007}^{+0.004}$ & $1.96_{-0.07}^{+0.06}$ \\
2b(i) & \qb & $0.068_{-0.007}^{+0.019}$ & $1.93_{-0.08}^{+0.20}$ \\
2b(i) & \qc & $0.058_{-0.009}^{+0.017}$ & $1.73_{-0.06}^{+0.19}$ \\
2b(i) & \qd & $0.088_{-0.012}^{+0.020}$ & $2.06_{-0.07}^{+0.12}$ \\

\enddata
\end{deluxetable}

\begin{deluxetable}{l c cc}
\tablewidth{\floatwidth}
\tablewidth{0pt}
\tablecaption{\label{tres-3dcorr}
Two-point correlation function $\xi_{2,3D}(r)$ for 3D cells
}
\tablehead{
\colhead{Sample} & \colhead{Quadrant}
 & \colhead{$r_0$($h^{-1}$ Mpc)} & \colhead{$\gamma$} 
}
\startdata
1a(g) & All & $5.64_{-0.02}^{+0.05}$ & $1.51_{-0.01}^{+0.01}$ \\
1a(g) & \qa & $5.74_{-0.12}^{+0.24}$ & $1.53_{-0.07}^{+0.04}$ \\
1a(g) & \qb & $5.72_{-0.32}^{+0.17}$ & $1.49_{-0.05}^{+0.12}$ \\
1a(g) & \qc & $5.27_{-0.12}^{+0.04}$ & $1.53_{-0.02}^{+0.06}$ \\
1a(g) & \qd & $6.14_{-0.41}^{+0.51}$ & $1.37_{-0.12}^{+0.09}$ \\
\\
1a(r) & All & $5.95_{-0.03}^{+0.05}$ & $1.51_{-0.01}^{+0.01}$ \\
1a(r) & \qa & $6.07_{-0.12}^{+0.24}$ & $1.54_{-0.07}^{+0.04}$ \\
1a(r) & \qb & $6.09_{-0.38}^{+0.18}$ & $1.49_{-0.05}^{+0.12}$ \\
1a(r) & \qc & $5.48_{-0.13}^{+0.04}$ & $1.54_{-0.02}^{+0.06}$ \\
1a(r) & \qd & $6.57_{-0.45}^{+0.55}$ & $1.38_{-0.10}^{+0.09}$ \\
\\
1a(i) & All & $6.05_{-0.03}^{+0.06}$ & $1.51_{-0.01}^{+0.01}$ \\
1a(i) & \qa & $6.19_{-0.13}^{+0.25}$ & $1.53_{-0.07}^{+0.03}$ \\
1a(i) & \qb & $6.16_{-0.39}^{+0.19}$ & $1.50_{-0.04}^{+0.13}$ \\
1a(i) & \qc & $5.55_{-0.14}^{+0.04}$ & $1.54_{-0.02}^{+0.06}$ \\
1a(i) & \qd & $6.73_{-0.45}^{+0.56}$ & $1.38_{-0.10}^{+0.09}$ \\
\\
\\
1b(g) & All & $7.78_{-0.10}^{+0.12}$ & $1.51_{-0.01}^{+0.02}$ \\
1b(g) & \qa & $8.19_{-0.45}^{+0.68}$ & $1.47_{-0.10}^{+0.09}$ \\
1b(g) & \qb & $7.19_{-0.64}^{+0.47}$ & $1.63_{-0.11}^{+0.27}$ \\
1b(g) & \qc & $7.25_{-0.54}^{+0.17}$ & $1.52_{-0.04}^{+0.10}$ \\
1b(g) & \qd & $7.75_{-0.83}^{+0.56}$ & $1.38_{-0.11}^{+0.11}$ \\
\\
1b(r) & All & $8.44_{-0.10}^{+0.13}$ & $1.57_{-0.01}^{+0.01}$ \\
1b(r) & \qa & $8.76_{-0.47}^{+0.84}$ & $1.53_{-0.10}^{+0.06}$ \\
1b(r) & \qb & $8.28_{-0.88}^{+0.67}$ & $1.61_{-0.08}^{+0.16}$ \\
1b(r) & \qc & $7.44_{-0.40}^{+0.19}$ & $1.63_{-0.03}^{+0.07}$ \\
1b(r) & \qd & $8.85_{-0.72}^{+1.17}$ & $1.49_{-0.22}^{+0.16}$ \\
\\
1b(i) & All & $8.44_{-0.10}^{+0.12}$ & $1.59_{-0.01}^{+0.01}$ \\
1b(i) & \qa & $8.38_{-0.42}^{+0.76}$ & $1.60_{-0.11}^{+0.07}$ \\
1b(i) & \qb & $8.38_{-0.73}^{+0.53}$ & $1.63_{-0.07}^{+0.13}$ \\
1b(i) & \qc & $7.70_{-0.48}^{+0.28}$ & $1.60_{-0.03}^{+0.07}$ \\
1b(i) & \qd & $9.09_{-0.75}^{+1.21}$ & $1.48_{-0.23}^{+0.14}$ \\
\\
\\
2b(g) & All & $7.23_{-0.05}^{+0.06}$ & $1.53_{-0.01}^{+0.01}$ \\
2b(g) & \qa & $7.07_{-0.22}^{+0.22}$ & $1.53_{-0.04}^{+0.04}$ \\
2b(g) & \qb & $7.02_{-0.10}^{+0.09}$ & $1.60_{-0.01}^{+0.02}$ \\
2b(g) & \qc & $7.32_{-0.18}^{+0.10}$ & $1.51_{-0.01}^{+0.04}$ \\
2b(g) & \qd & $7.34_{-0.61}^{+0.71}$ & $1.53_{-0.08}^{+0.11}$ \\
\\
2b(r) & All & $7.83_{-0.06}^{+0.06}$ & $1.54_{-0.01}^{+0.01}$ \\
2b(r) & \qa & $7.60_{-0.23}^{+0.23}$ & $1.54_{-0.04}^{+0.04}$ \\
2b(r) & \qb & $7.71_{-0.16}^{+0.13}$ & $1.59_{-0.01}^{+0.02}$ \\
2b(r) & \qc & $7.89_{-0.19}^{+0.08}$ & $1.55_{-0.01}^{+0.05}$ \\
2b(r) & \qd & $7.94_{-0.76}^{+0.76}$ & $1.53_{-0.08}^{+0.12}$ \\
\\
2b(i) & All & $7.98_{-0.06}^{+0.06}$ & $1.55_{-0.01}^{+0.01}$ \\
2b(i) & \qa & $7.73_{-0.22}^{+0.23}$ & $1.54_{-0.04}^{+0.03}$ \\
2b(i) & \qb & $7.92_{-0.16}^{+0.13}$ & $1.58_{-0.01}^{+0.02}$ \\
2b(i) & \qc & $8.02_{-0.18}^{+0.06}$ & $1.56_{-0.01}^{+0.05}$ \\
2b(i) & \qd & $8.14_{-0.85}^{+0.86}$ & $1.53_{-0.11}^{+0.16}$ \\
\enddata
\end{deluxetable}

Since previous studies~(e.g. \citealt{2003MNRAS.346...78H,2002ApJ...579...42C}) have shown that the two-point correlation function deviates from a power law at large scales, we use datapoints from scales smaller than $10.0 h^{-1}$ Mpc for 3D cells or scales smaller than $1.25^{\circ}$ for 2D cells to determine a power law fit. Because $\xi_2(r)$ depends on the gradient of $V\overline{\xi}_2(r)$, we use a 5-point moving average of $V\overline{\xi}_2(r)$ to obtain the overall shape of $V\overline{\xi}_2(r)$ for our analysis. We summarize our results in tables \ref{tres-2dcorr} and \ref{tres-3dcorr}.

We find that the value of $\gamma$ for the 3D two-point correlation function is about $1.5\sim1.6$ and $\xi_{2,3D}(r)$ begins to deviate from a power law at scales of $r = 11\sim12 h^{-1}$ Mpc, in agreement with earlier work such as \citet{2003MNRAS.346...78H}. The value of $\gamma$ for the 2D two-point correlation function is about $1.7\sim1.8$ and is close to the value obtained by \citet{2002ApJ...579...42C}. We note that $\xi_{2,2D}(\theta)$ shows a break from the power law fit at $\theta \approx 1.75^\circ$ for the 1a and 2b samples, and $\theta \approx 2.4^\circ \sim 2.6^\circ$ for the 1b samples.

To compare the 2D and 3D samples, we first note that the 2D samples measure the projected correlation function while the 3D samples measure the redshift space correlation function. The 3D samples are affected by distortions in redshift space caused by peculiar velocities while the 2D samples, which ignore the detailed distance information, are not affected by redshift space distortions. Therefore the comparison between 2D and 3D samples can help quantify the effect of these distortions.

\citet{1969PASJ...21..221T} and \citet{1983ApJ...267..465D} relate the projected correlation function $\xi_{2,2D}(r_p)$ to the real space corelation function $\xi_2(r_{real})$ by
\begin{equation}\label{eres-wxi}
\xi_2(r_{real}) = -\frac{1}{r_r}\int_{r_{real}}^\infty \frac{d\xi_{2,2D}}{dr_p}\left(r_p^2-r_{real}^2\right)^{-1/2}dr_p
\end{equation}
where $r_{real}$ is the real space position, $r_r$ is the radial position along the line of sight and $r_p$ is the position in the direction perpendicular to the line of sight. Because we write the projected two-point correlation function in terms of angles on the sky, equation \refeq{eres-wxi} does not directly apply to the comparison between the 2D and 3D samples in this study without a conversion factor between $\theta$ and $r_p$. However, a detailed analysis of equation \refeq{eres-wxi} by \citet{1969PASJ...21..221T} shows that for a power law where $\xi_{2,2D}(\theta) \propto \theta^{-\gamma+1}$, the real space correlation function follows $\xi_2(r_{real}) \propto r_{real}^{-\gamma}$. This means that the value of $\gamma$ for the 2D projected samples follows the value of $\gamma$ for the correlation function in real space. Hence we can simply compare the value of $\gamma$ between the 2D projected samples and 3D redshift space samples to quantify the redshift space distortions.

Comparisons of tables \ref{tres-2dcorr} and \ref{tres-3dcorr} shows that the value of $\gamma$ for the 2D samples~(projection) is consistently higher than the value of $\gamma$ for the 3D samples~(redshift space) with a difference of about $0.2 \sim 0.3$, agreeing with earlier work by \citet{1994ApJ...425....1F} and \citet{2003MNRAS.346...78H}. These agreements with earlier work show that the counts-in-cells method used in this study correctly describes galaxy clustering and distortions in redshift space. However, because distortions in redshift space will affect an analysis of large-scale structure using the 3D samples, we note that results from the 3D samples may be less reliable than the results from the 2D samples.

\begin{figure*}
\begin{center}
\includegraphics[width=\floatwidth]{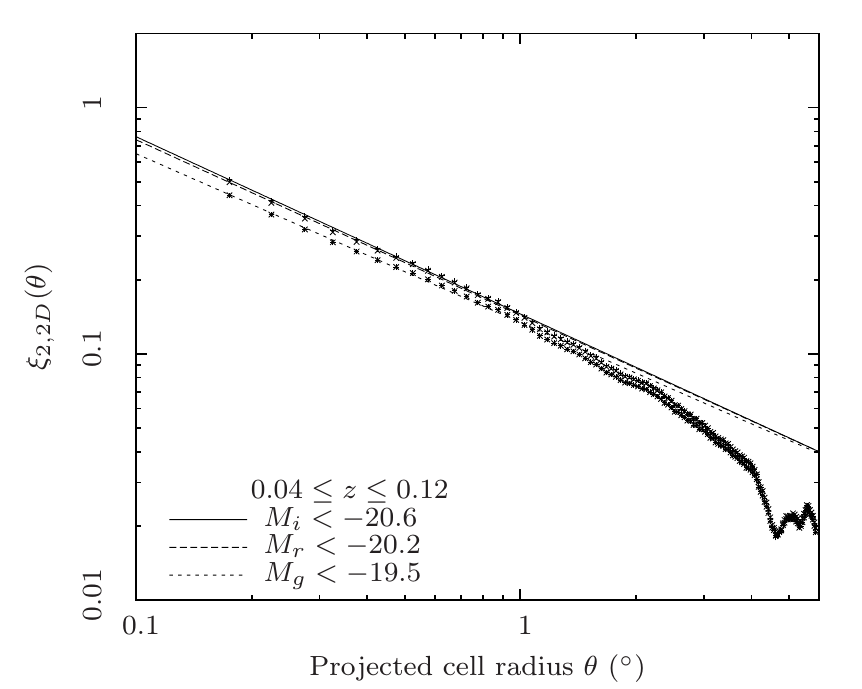}
\includegraphics[width=\floatwidth]{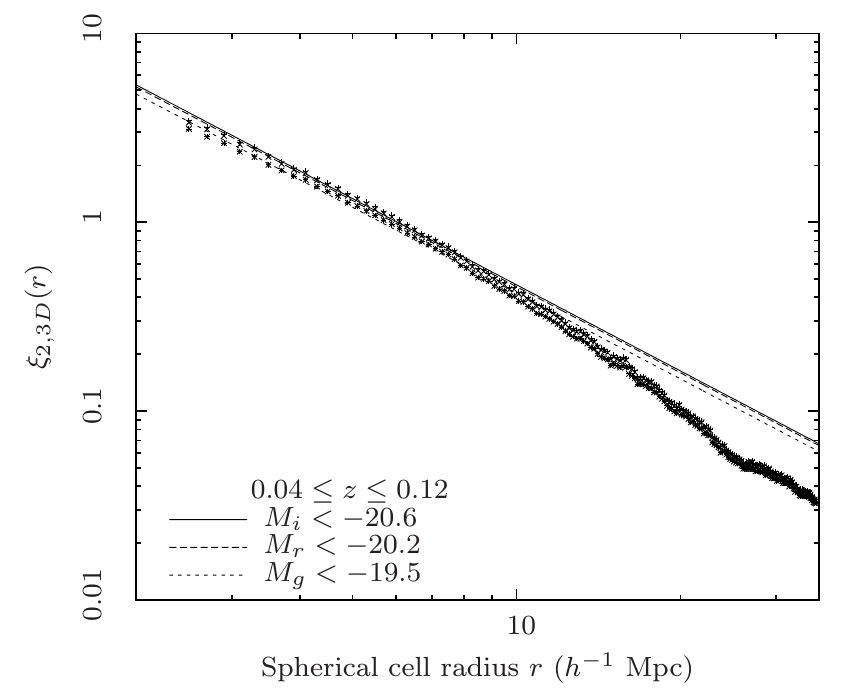}\\
\includegraphics[width=\floatwidth]{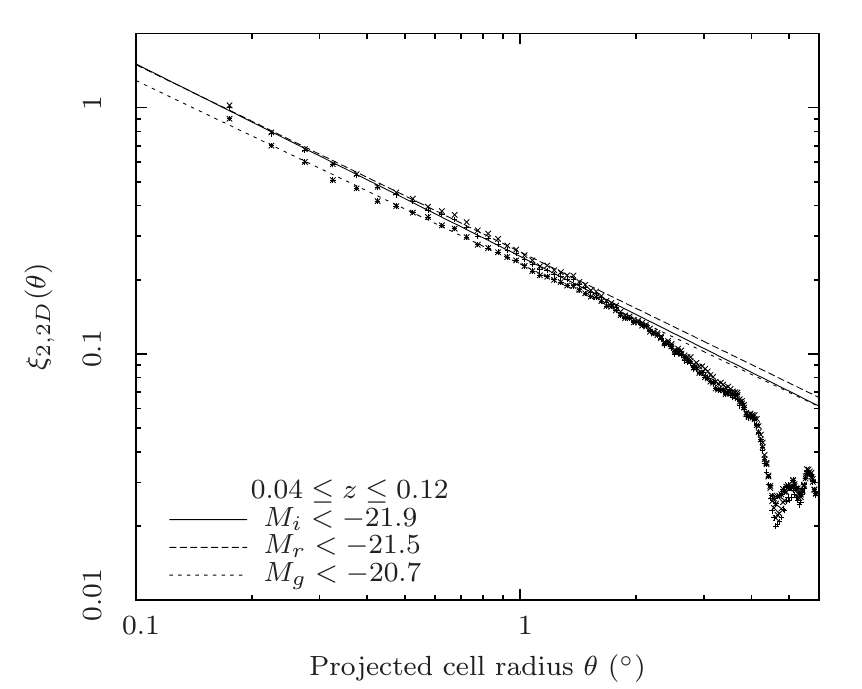}
\includegraphics[width=\floatwidth]{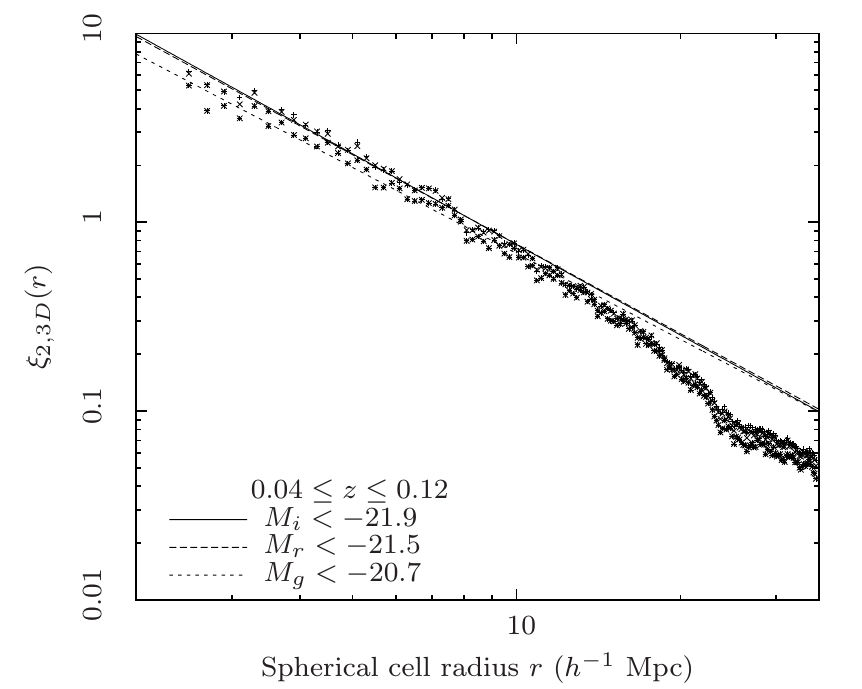}\\
\includegraphics[width=\floatwidth]{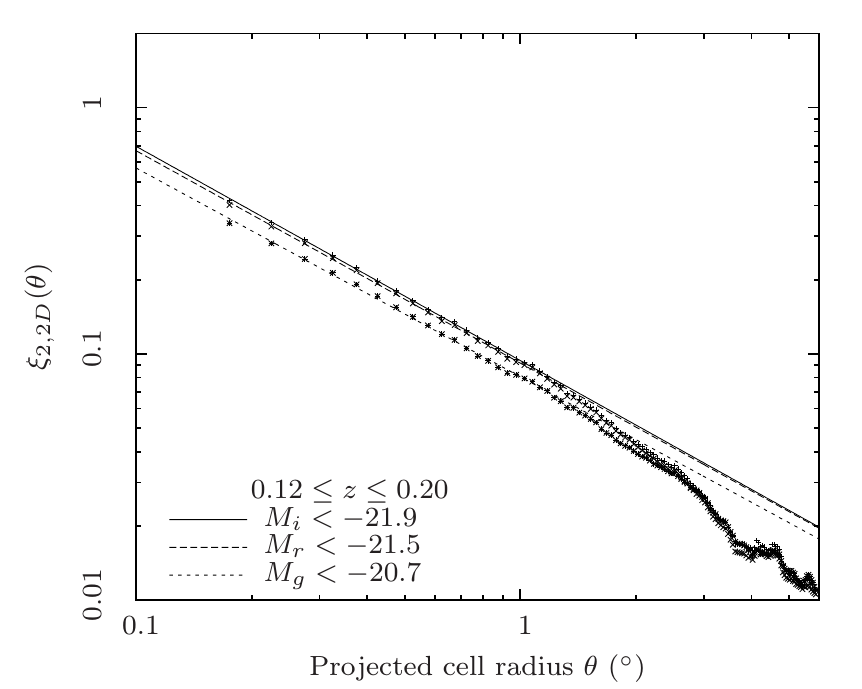}
\includegraphics[width=\floatwidth]{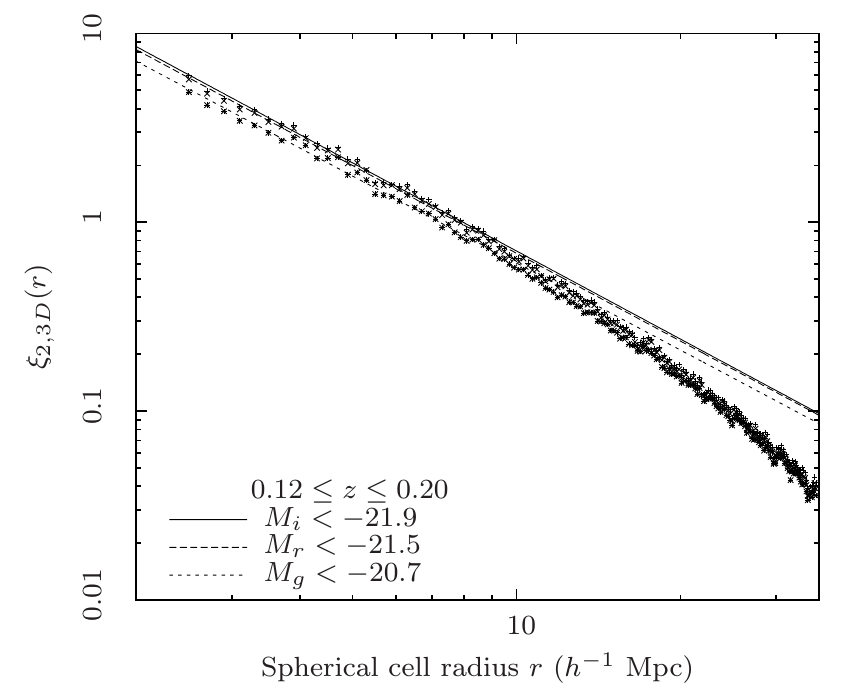}
\end{center}
\caption{\label{fres-2ptc}
Two-point correlation functions of different samples. Top left: 2D cells, 1a samples. Middle left: 2D cells, 1b samples. Bottom left: 2D cells, 2b samples. Top right: 3D cells, 1a samples. Middle right: 3D cells,1b samples. Bottom right: 3D cells, 2b samples. 
}
\end{figure*}

Comparing samples with different color selection criteria, we note that there is generally no significant difference in the value of $\gamma$ between samples selected using different colors within the same redshift range although in some cases there is a difference in $r_0$ and $\theta_0$ which can be attributed to the fact that samples selected using different colors have slightly different spatial densities~(c.f. table \ref{tdata-zcut}). This means that samples selected using different colors cluster similarly. For this reason, we will focus our subsequent analysis on the $r$-band selected samples because they have the highest spatial density. We illustrate the comparison between the selection criteria for different colors in figure \ref{fres-2ptc} and note in particular that the slopes of the power law fits for samples selected using different colors are close to each other.

Comparing the power law fit parameters $r_0$, $\theta_0$ and $\gamma$ between quadrants, we note the presence of differences that are significant at the 1-$\sigma$ level. In particular, these differences are consistent across different color and magnitude cuts but not across redshift ranges. For example, the value of $\gamma$ in the q2 subsample for 2D cells is significantly lower than the other quadrants for the low redshift range, but this is not the case in the high redshift range. We see similar behavior for the q3 subsample in the 2D high redshift range, and for the q4 subsample in the 3D low redshift range. For more insight into the differences between quadrants, we look at the counts-in-cells distribution $f_V(N)$.

\subsection{Counts-in-cells distribution $f_V(N)$}

Using measurements of the two-point correlation function, we can define 3 regimes to examine in detail. Since the two-point correlation function exhibits a break at about $12 h^{-1}$ Mpc for 3D cells and at about $2^\circ$ for 2D cells, we look at $f_V(N)$ for the cell sizes of $6.0 h^{-1}$ Mpc, $12.0 h^{-1}$ Mpc and $24.0 h^{-1}$ Mpc for 3D cells, and $1.00^\circ$, $2.00^\circ$ and $4.00^\circ$ for 2D cells. Because the analysis of the two-point correlation has shown samples selected using different colors are similarly clustered, we focus our subsequent analysis of the counts-in-cells distribution on the $r$-band samples because they have the highest spatial density for a given redshift and magnitude cut. We summarize the cells we use in table \ref{tres-cellsize}. 

\begin{deluxetable}{lcc}
\tablewidth{\floatwidth}
\tablewidth{0pt}
\tablecaption{\label{tres-cellsize}
Cells used for counts-in-cells analysis
}
\tablehead{
\colhead{Size} & \colhead{Redshift} & \colhead{Comoving Volume}
\\
 & & ($10^4 h^{-3}$ Mpc)
}
\startdata
\hline
$\theta$ ($^\circ$) & \multicolumn{2}{l}{2D Projected Cells} \\
\hline
$1.00$ & $0.04 \leq z \leq 0.12$ & $\phn1.31$ \\
$2.00$ & $0.04 \leq z \leq 0.12$ & $\phn5.25$ \\
$4.00$ & $0.04 \leq z \leq 0.12$ & $20.98$ \\
\\
$1.00$ & $0.12 \leq z \leq 0.20$ & $\phn4.60$ \\
$2.00$ & $0.12 \leq z \leq 0.20$ & $18.38$ \\
$4.00$ & $0.12 \leq z \leq 0.20$ & $73.49$ \\

\hline
$r$ ($h^{-1}$ Mpc) & \multicolumn{2}{l}{3D Spherical Cells} \\
\hline
$ 6.0$ & $0.04 \leq z \leq 0.12$ & $0.090$ \\
$12.0$ & $0.04 \leq z \leq 0.12$ & $0.724$ \\
$24.0$ & $0.04 \leq z \leq 0.12$ & $5.79\phn$ \\
\\
$ 6.0$ & $0.12 \leq z \leq 0.20$ & $0.090$ \\
$12.0$ & $0.12 \leq z \leq 0.20$ & $0.724$ \\
$24.0$ & $0.12 \leq z \leq 0.20$ & $5.79\phn$ \\
\enddata
\end{deluxetable}

From the analysis of the two-point correlation function, we note that there may be considerable variation between quadrants and hence the sample we have used might not be homogenous. A simple test to show that the full sample is homogenous is to show that different quadrants, being disjoint subsamples of the full sample, are identically distributed. To compare quadrants, we use a random permutation test~\citep{1957TAMS-28-181} with $100,000$ permutations to compare the equivalence of $\overline{N}$ and $\langle(\Delta N)^2\rangle$ across quadrants. This gives a necessary condition for the distribution of two samples to be equivalent.

To compare $\overline{N}$, we define the observed test statistic $T_N(obs) = |\overline{N}_A - \overline{N}_B|$ as the difference in $\overline{N}$ between two given quadrants A and B, the test statistic $T_N = |\overline{N}_Ap - \overline{N}_Bp|$ as the difference in $\overline{N}$ for a given permutation $p$, and the null hypothesis that both quadrants have the same $\overline{N}$. Each permutation is constructed by shuffling cells at random across quadrants. The p-value is given by the frequency of sampled permutations where the $T_N \geq T_N(obs)$. We can define a similar test statistic and procedure $T_\Delta = |\langle(\Delta N)^2\rangle_Ap - \langle(\Delta N)^2\rangle_Bp|$ for the variance.

We find that at the 95\% level, in the 3D 1a(r) $6.0 h^{-1}$ Mpc sample, q1 and q4 have the same mean and variance. In the 2b(r) $6.0 h^{-1}$ Mpc sample, the q1-q2, q1-q3 and q3-q4 pairs have the same mean and variance, and in the 2b(r) $12.0 h^{-1}$ Mpc sample, the q1-q2 and q2-q3 pairs have the same mean and variance. In all other samples, we find that we can reject the null hypothesis that $\overline{N}$ and $\langle(\Delta N)^2\rangle$ are equal in the samples at the 95\% level. Here we note that the q1-q4 and q2-q4 pairs in the 2b(r) $6.0 h^{-1}$ Mpc sample and the q2-q3 pair in the 2b(r) $12.0 h^{-1}$ Mpc sample was not found to have the same mean and variance, suggesting that although there are quadrant pairs that seem to be similar, there is still enough variation between 3 quadrants such that not every pair of quadrants is identically distributed.

\begin{figure*}
\begin{center}
\includegraphics[width=\floatwidth]{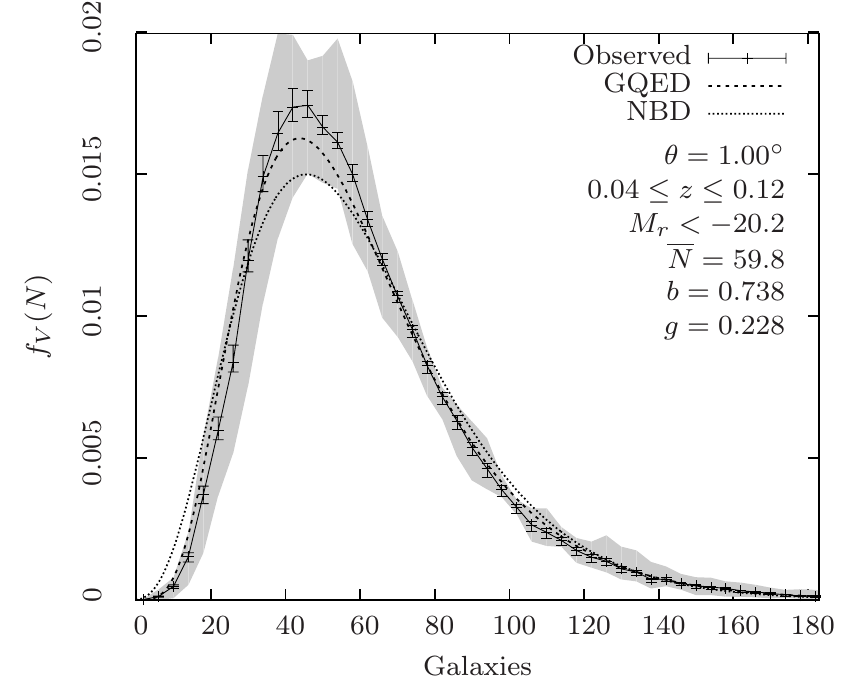}
\includegraphics[width=\floatwidth]{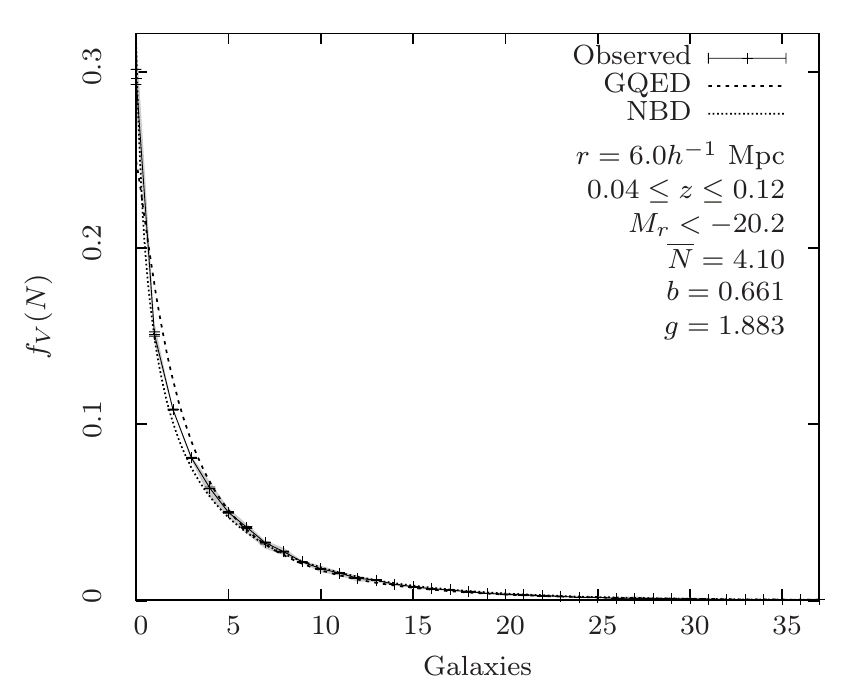}\\
\includegraphics[width=\floatwidth]{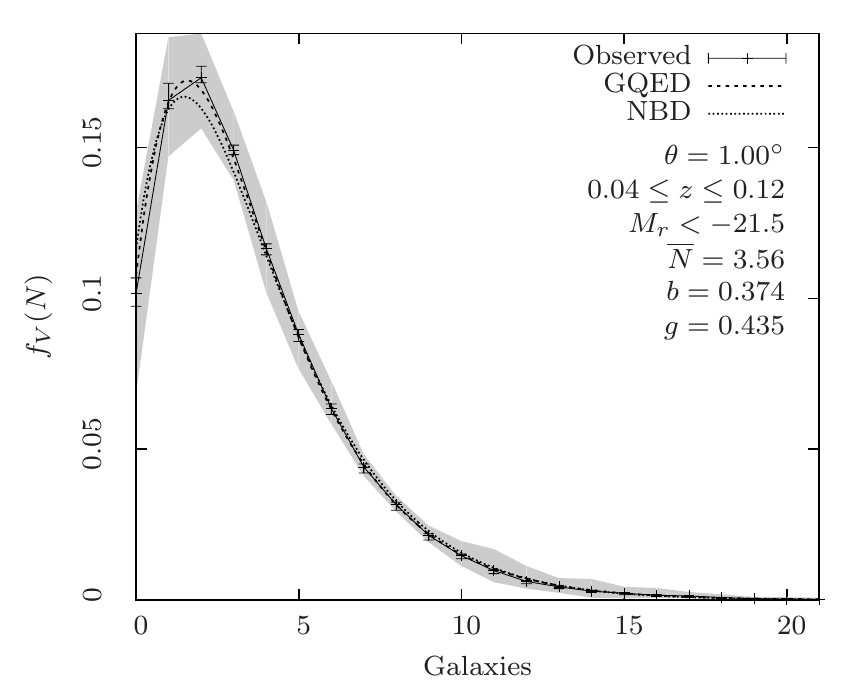}
\includegraphics[width=\floatwidth]{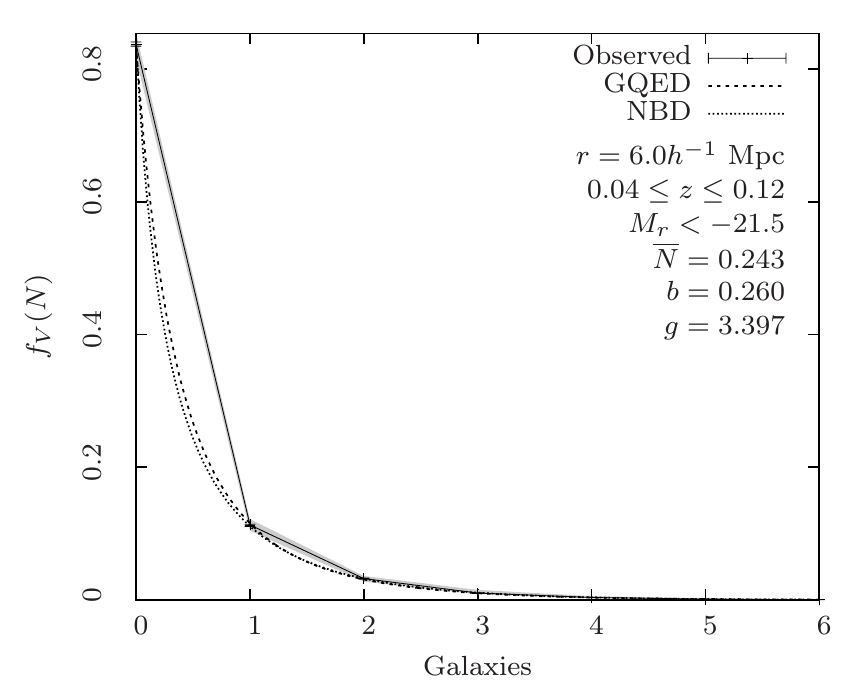}\\
\includegraphics[width=\floatwidth]{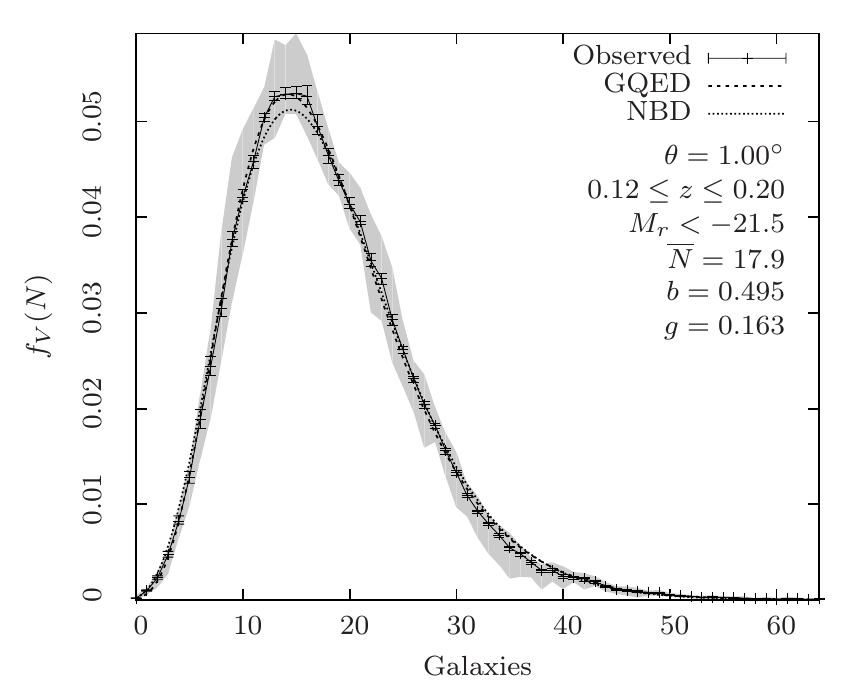}
\includegraphics[width=\floatwidth]{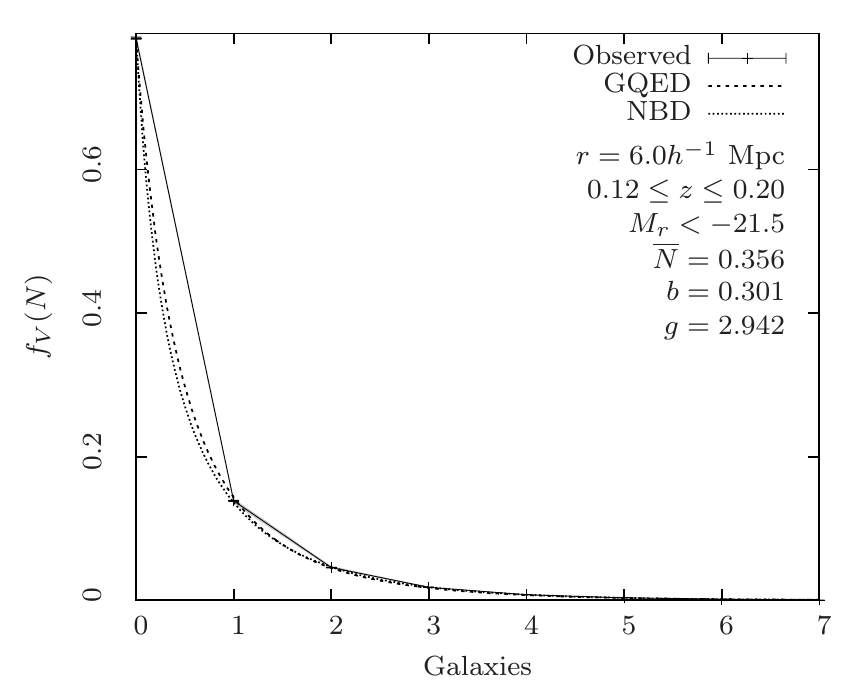}
\end{center}
\caption{\label{fres-fvnc1}
$r$-band counts-in-cells of different samples. Top left: $1.00^\circ$ cells, 1a(r) sample; Middle left: $1.00^\circ$ cells, 1b(r) sample; Bottom left: $1.00^\circ$ cells, 2b(r) sample. Top right: $6.0 h^{-1}$ Mpc cells, 1a(r) sample; Middle right: $6.0 h^{-1}$ Mpc cells, 1b(r) sample; Bottom right: $6.0 h^{-1}$ Mpc cells, 2b(r) sample. The shaded band represents the extent of quadrant to quadrant variations and the errorbars represent the jackknife errors.
}
\end{figure*}

\begin{figure*}
\begin{center}
\includegraphics[width=\floatwidth]{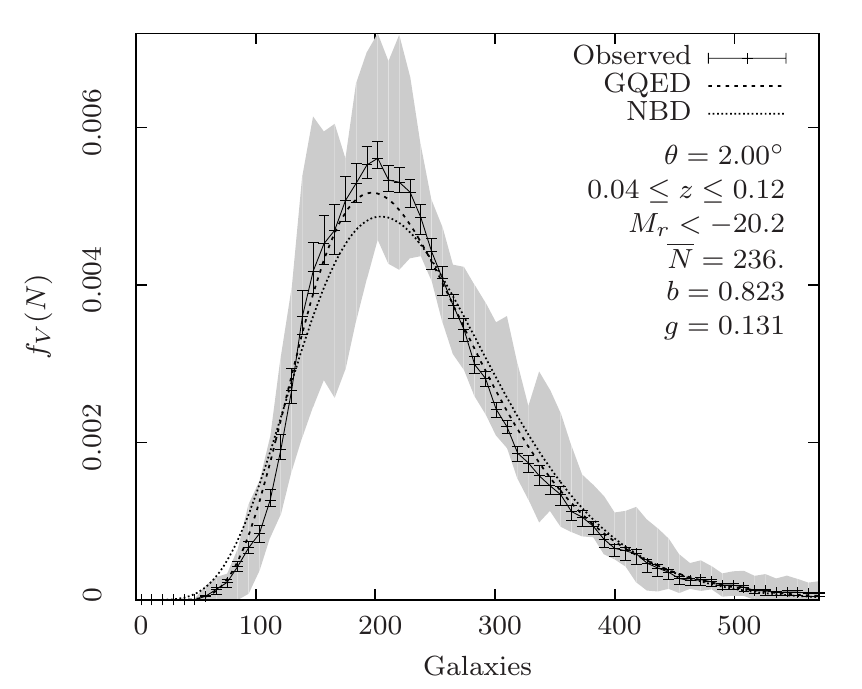}
\includegraphics[width=\floatwidth]{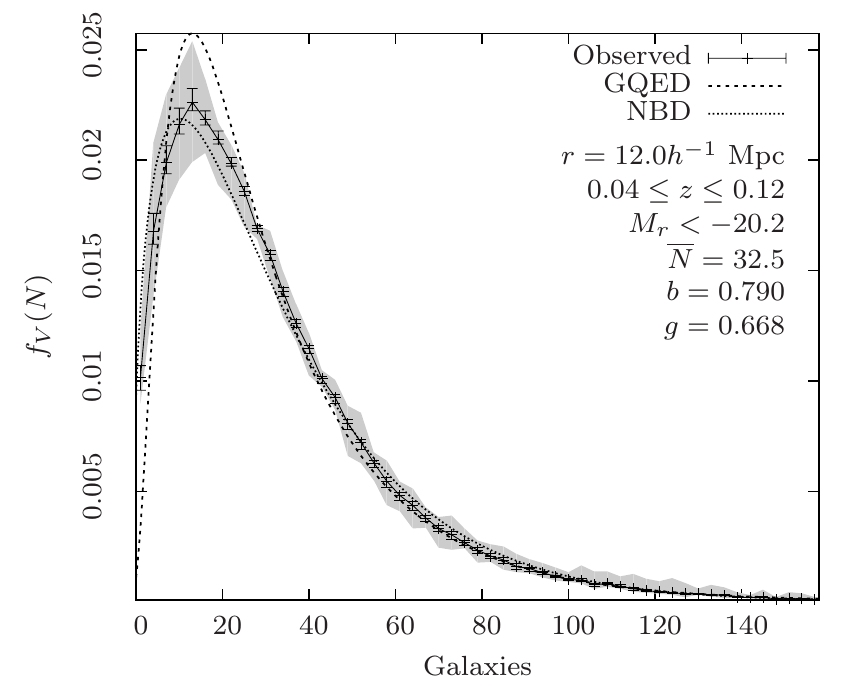}\\
\includegraphics[width=\floatwidth]{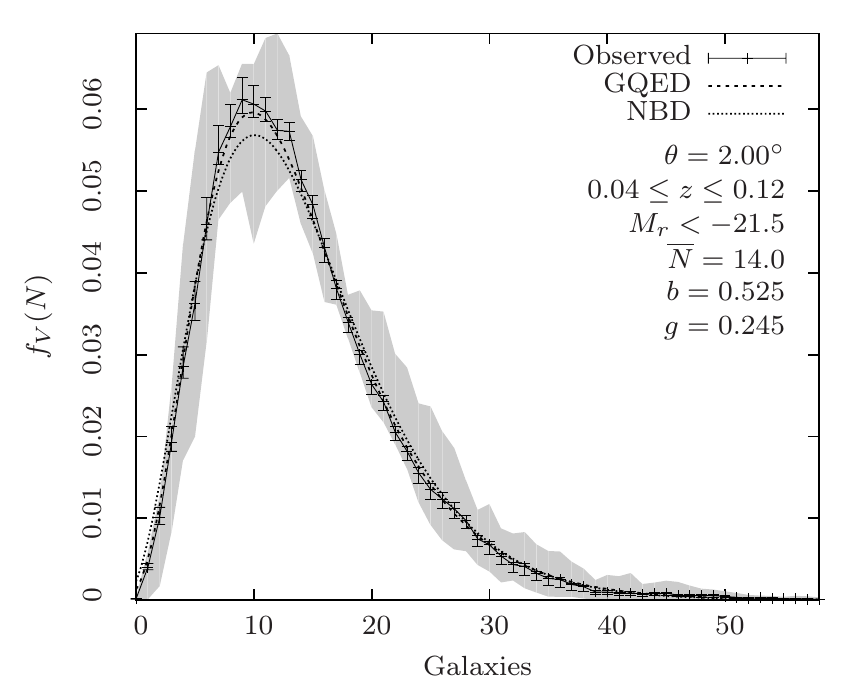}
\includegraphics[width=\floatwidth]{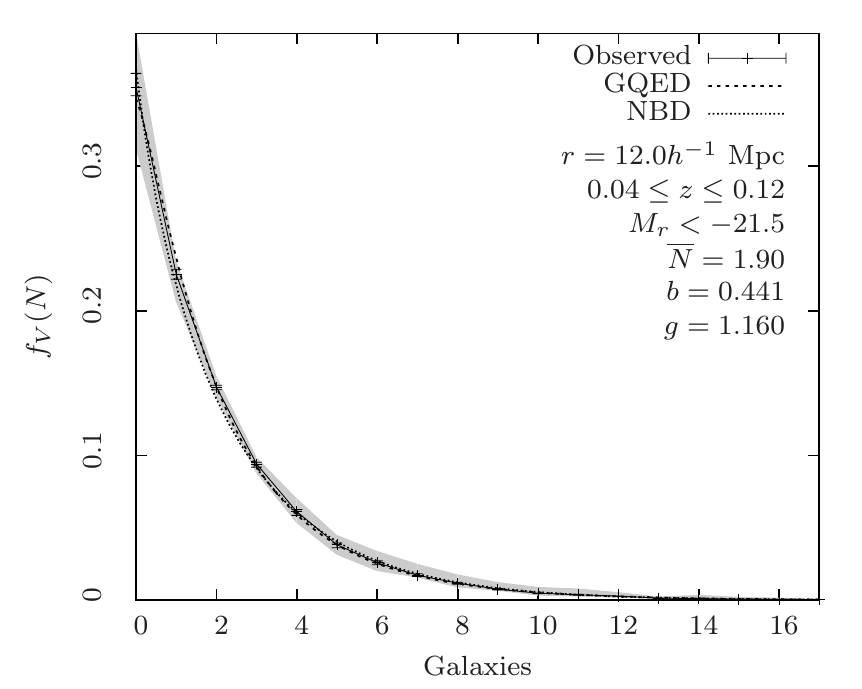}\\
\includegraphics[width=\floatwidth]{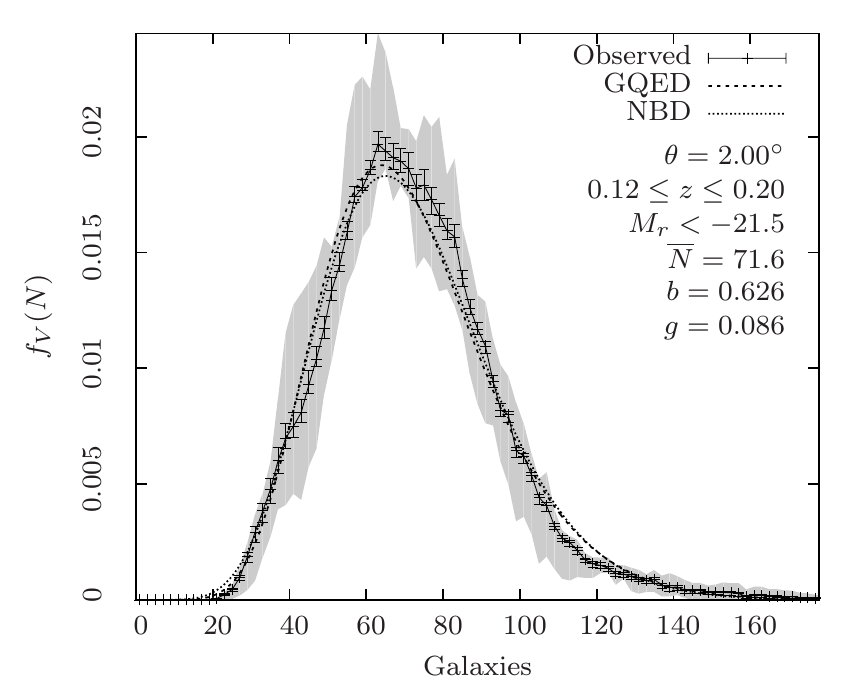}
\includegraphics[width=\floatwidth]{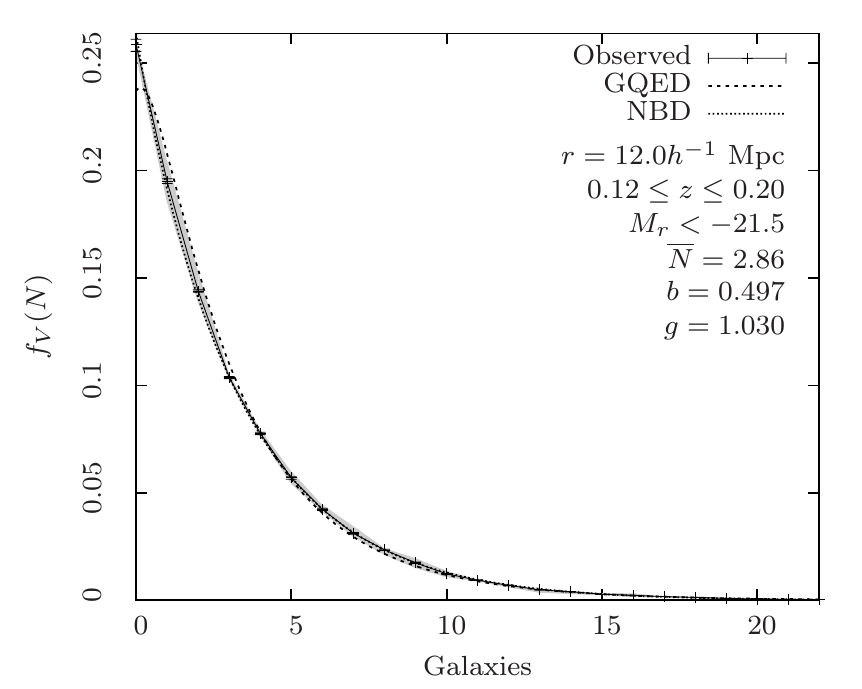}
\end{center}
\caption{\label{fres-fvnc2}
$r$-band counts-in-cells of different samples. Top left: $2.00^\circ$ cells, 1a(r) sample; Middle left: $2.00^\circ$ cells, 1b(r) sample; Bottom left: $2.00^\circ$ cells, 2b(r) sample. Top right: $12.0 h^{-1}$ Mpc cells, 1a(r) sample; Middle right: $12.0 h^{-1}$ Mpc cells, 1b(r) sample; Bottom right: $12.0 h^{-1}$ Mpc cells, 2b(r) sample. The shaded band represents the extent of quadrant to quadrant variations and the errorbars represent the jackknife errors.
}
\end{figure*}

\begin{figure*}
\begin{center}
\includegraphics[width=\floatwidth]{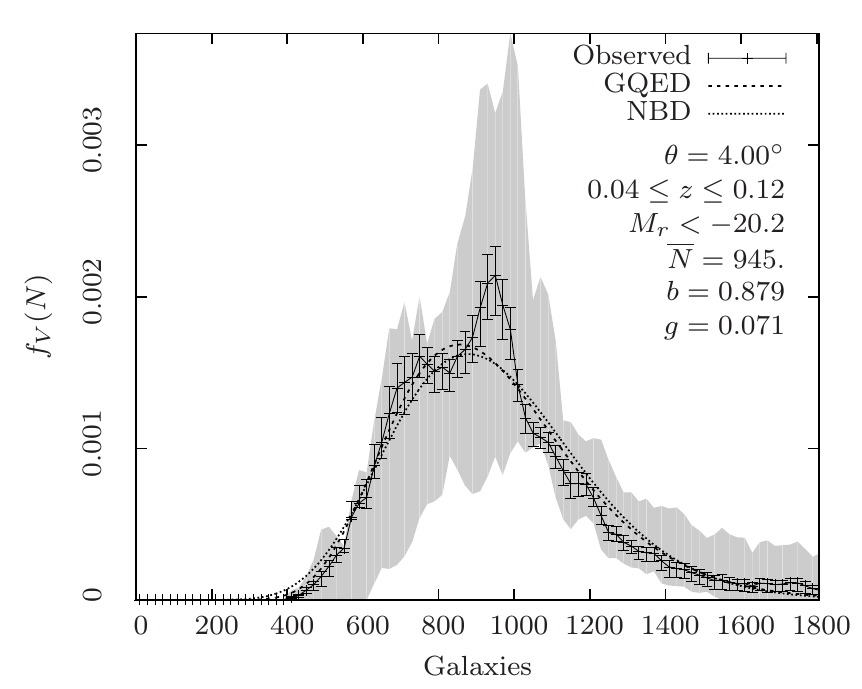}
\includegraphics[width=\floatwidth]{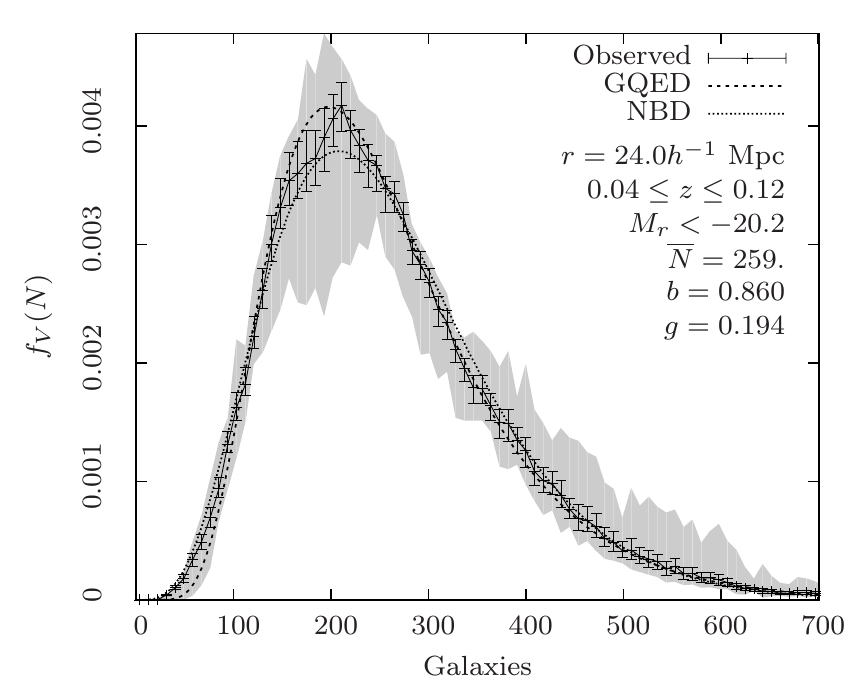}\\
\includegraphics[width=\floatwidth]{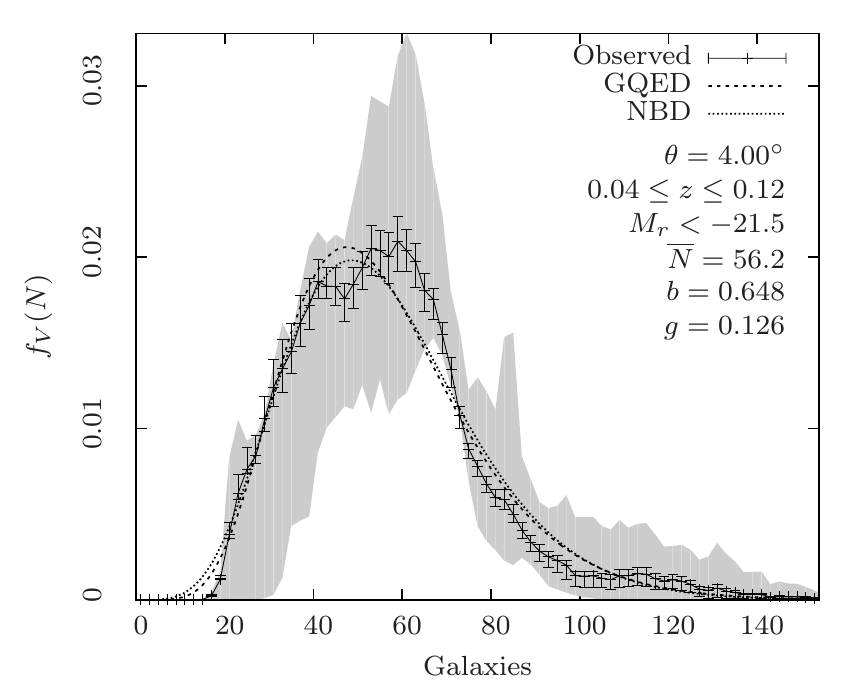}
\includegraphics[width=\floatwidth]{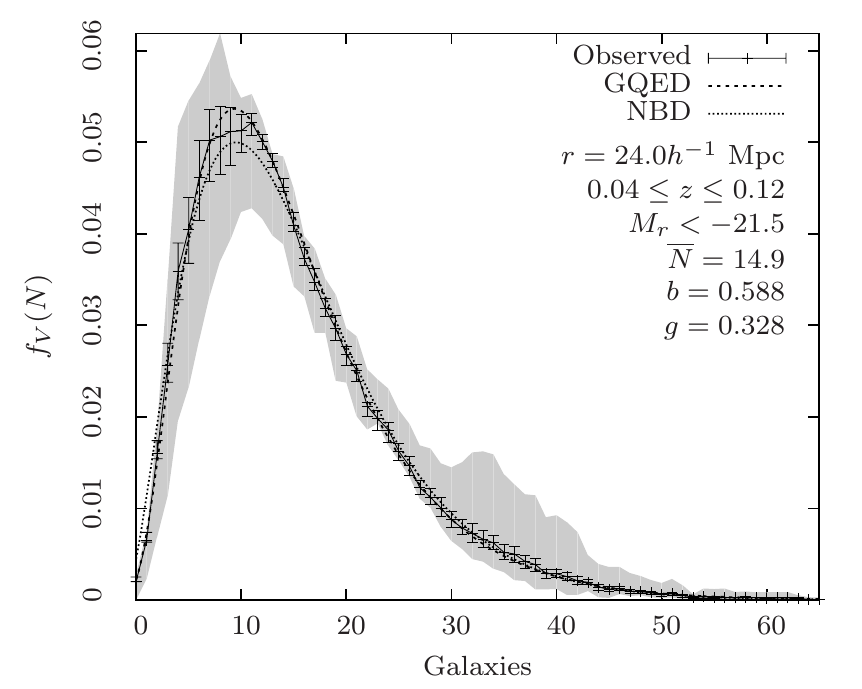}\\
\includegraphics[width=\floatwidth]{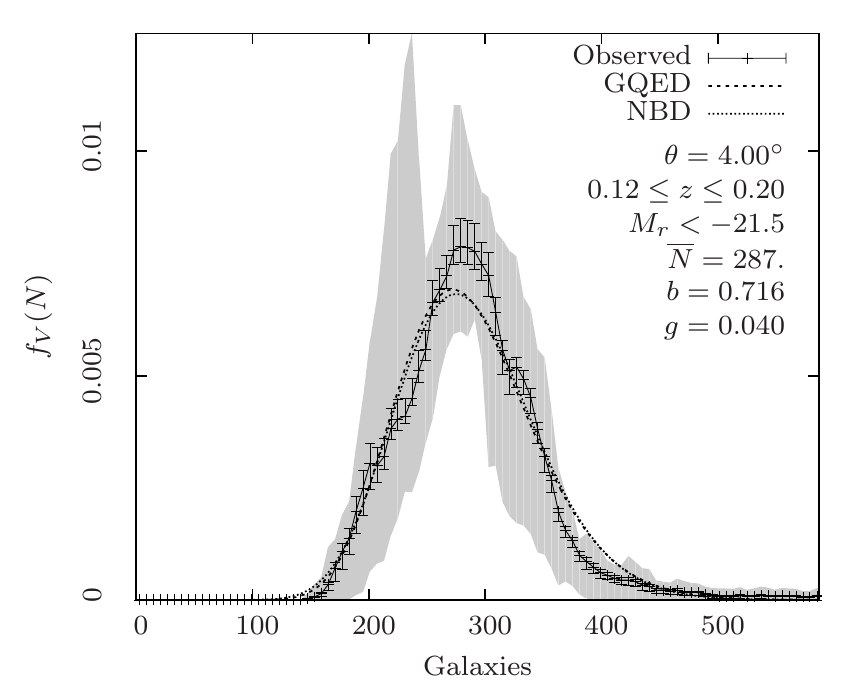}
\includegraphics[width=\floatwidth]{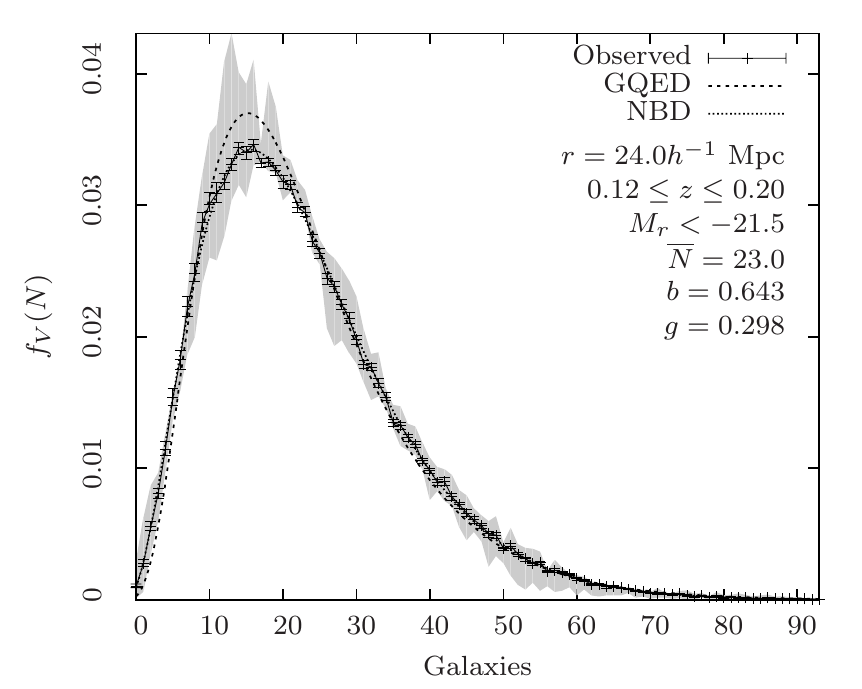}
\end{center}
\caption{\label{fres-fvnc3}
$r$-band counts-in-cells of different samples. Top left: $4.00^\circ$ cells, 1a(r) sample; Middle left: $4.00^\circ$ cells, 1b(r) sample; Bottom left: $4.00^\circ$ cells, 2b(r) sample. Top right: $24.0 h^{-1}$ Mpc cells, 1a(r) sample; Middle right: $24.0 h^{-1}$ Mpc cells, 1b(r) sample; Bottom right: $24.0 h^{-1}$ Mpc cells, 2b(r) sample. The shaded band represents the extent of quadrant to quadrant variations and the errorbars represent the jackknife errors.
}
\end{figure*}

The result that quadrants are generally not identically distributed suggests that the jackknife errors underestimate the true range of variability in the data. For this reason, a better and simpler estimate of variability would be to compare $f_V(N)$ across quadrants. Hence we use the quadrant to quadrant variations in $f_V(N)$ as a measure of the variability of the counts-in-cells statistic. We plot the observed counts-in-cells distribution $f_V(N)$, the minimum and maximum values of $f_V(N)$ from each quadrant as a shaded band, and the GQED and NBD with the same mean and variance as the observed $f_V(N)$ in figures \ref{fres-fvnc1}, \ref{fres-fvnc2} and \ref{fres-fvnc3}. We also plot the jackknife errors as errorbars to illustrate the difference between the jackknife errors and the quadrant to quadrant variations.

\begin{figure*}
\begin{center}
\includegraphics[width=\floatwidth]{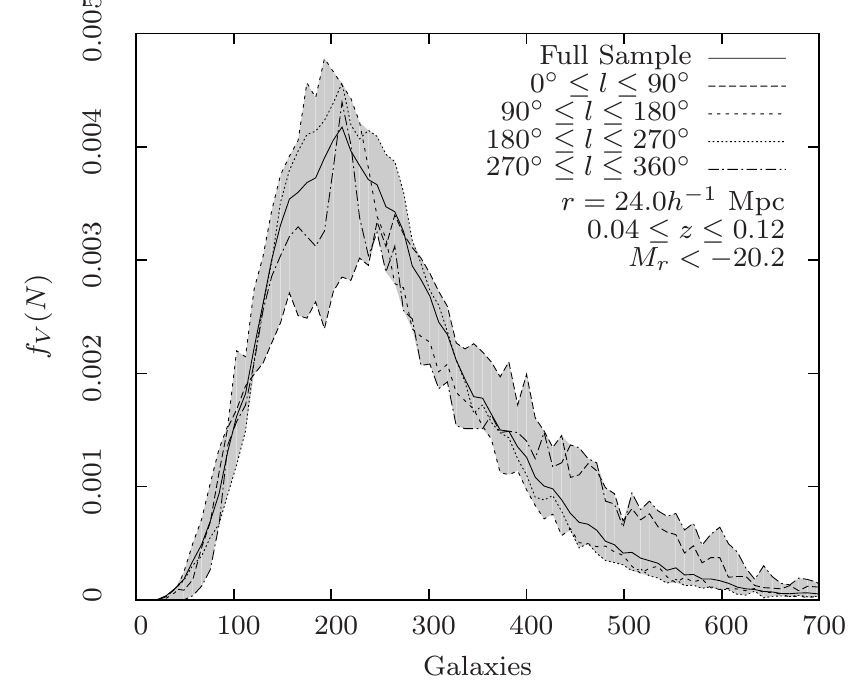}
\includegraphics[width=\floatwidth]{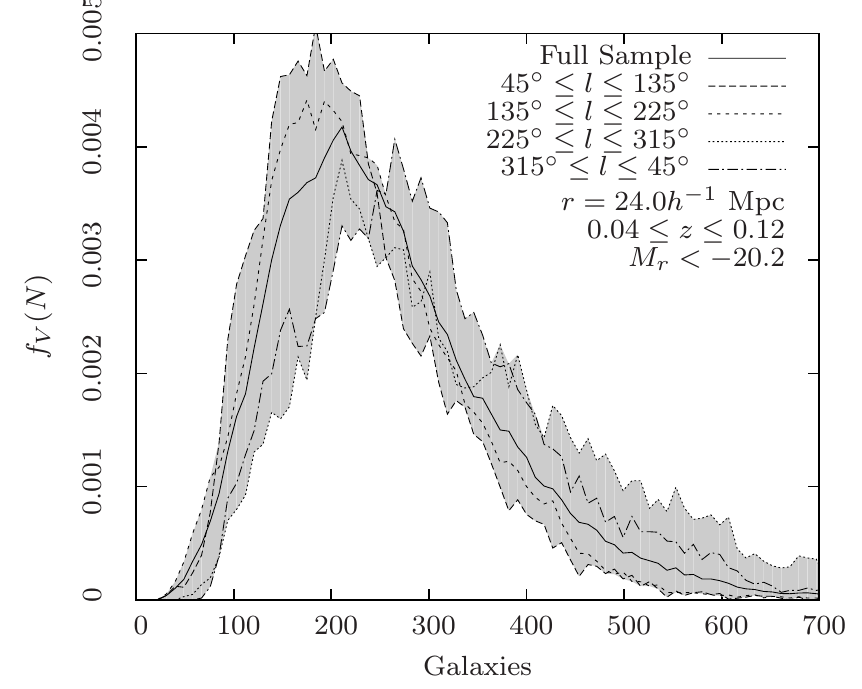}\\
\includegraphics[width=\floatwidth]{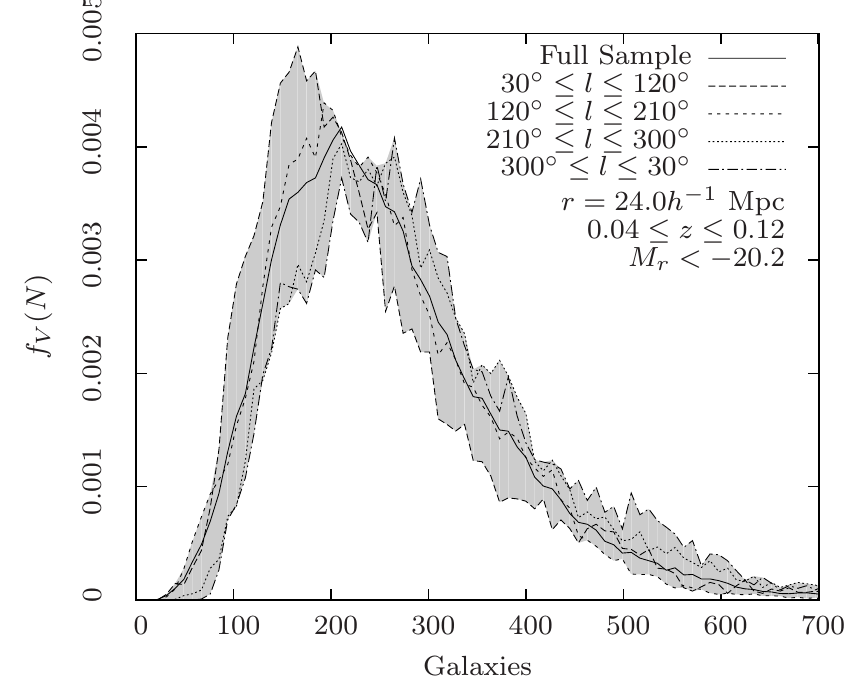}
\includegraphics[width=\floatwidth]{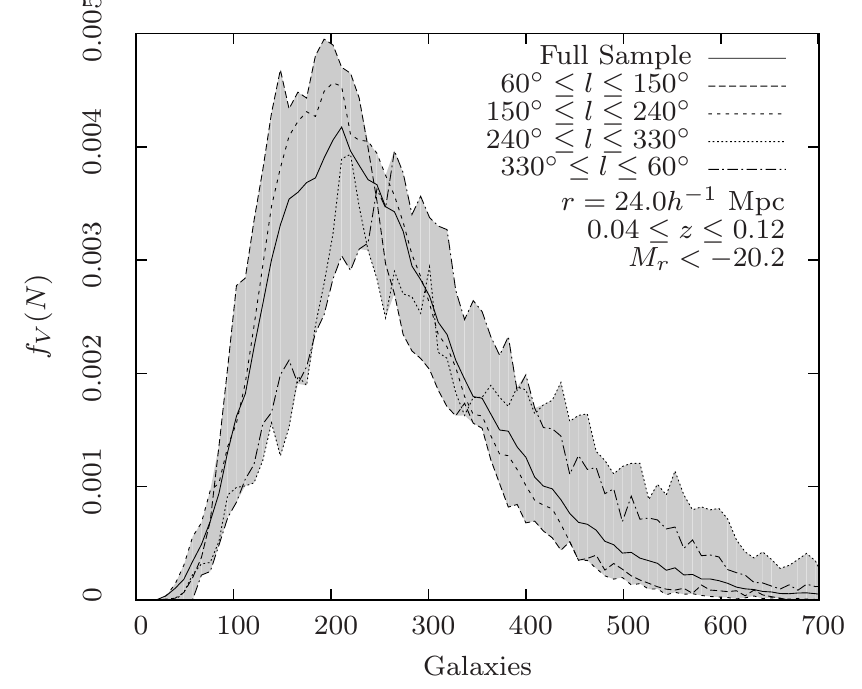}
\end{center}
\caption{\label{fres-fvnquad}
$r$-band counts-in-cells for $24.0 h^{-1}$ Mpc cells with different quadrant boundaries. Top left: $0^\circ$ shift; Top right: $45^\circ$ shift; Bottom left: $30^\circ$ shift; Bottom right: $60^\circ$ shift. The shaded band represents the extent of quadrant to quadrant variations using the selected quadrant boundaries. For clarity, errorbars are not plotted here.
}
\end{figure*}

We also repeat the analysis for different quadrant boundaries shifted by $30^\circ$, $45^\circ$ and $60^\circ$ in galactic longitude. In all cases, we see a similar amount of variation which suggests that in general, the amount of variation between quadrants is not coincidental with the choice of quadrant boundaries. To illustrate, we plot the variation between quadrants for different quadrant boundaries in figure \ref{fres-fvnquad}. We note from figure \ref{fres-fvnquad} that although the variation between quadrants is different for different quadrant boundaries, the amount of variation is approximately the same. This means that these variations are probably caused by fluctuations in the number density of galaxies on scales that are about as large as the quadrants themselves, suggesting a significant amount of cosmic variance.

\subsubsection{Comparison with models}
To study the parameters $b$ and $g$ and compare the observed $f_V(N)$ to models, we use equations \refeq{edist-bxi} and \refeq{edist-gxi} to obtain $b$ and $g$ in a self-consistent manner from the mean $\overline{N}$ and variance $\langle (\Delta N)^2\rangle$ such that the theoretical distributions have the same mean and variance as the observed $f_V(N)$. Since the populations of nearby cells are often strongly correlated, the cells are in general not independent and a $\chi^2$ fit cannot be used. Instead, we compute the least squares distance
\begin{equation} \label{eres-lsqdist}
x = \sum_{N=0}^{N_{max}} \left(f_V(N)_{obs} - f_V(N)\right)^2
\end{equation}
between the observed distribution and the theoretical distribution as a qualitative measure of goodness of fit where $N_{max}$ is the largest number of galaxies in a cell. We use this goodness-of-fit measure to determine which model is closer to the observed data.

\begin{deluxetable*}{lc rccc cc}
\tablewidth{0pt}
\tablecaption{\label{tres-2dfvn}
$r$-band 2D Counts-in-cells $f_{V}(N)$
}
\tablehead{
\colhead{Sample} & \colhead{Quadrant} & \colhead{Cells} & \colhead{$\overline{N}$} & \colhead{$b$} & \colhead{$g$} & \colhead{$x\times10^{-5}$} & \colhead{$x\times10^{-5}$} \\
 & & & & & & \colhead{(GQED)} & \colhead{(NBD)}
}
\startdata
\hline
\multicolumn{8}{c}{$\theta = 1.00^\circ$} \\
\hline
1a(r) & All & $135995$ & $59.8$ & $0.738$ & $0.228$ & $\phn5.68$ & $19.4$ \\
1a(r) & \qa &  $35174$ & $64.8$ & $0.773$ & $0.285$ & $10.3\phn$ & $27.4$ \\
1a(r) & \qb &  $31916$ & $54.2$ & $0.718$ & $0.213$ & $\phn9.94$ & $21.3$ \\
1a(r) & \qc &  $45518$ & $59.0$ & $0.706$ & $0.179$ & $\phn6.60$ & $16.9$ \\
1a(r) & \qd &  $15710$ & $64.7$ & $0.728$ & $0.193$ & $32.6\phn$ & $51.4$ \\
\\
1b(r) & All & $135995$ & $3.56$ & $0.374$ & $0.435$ & $\phn6.44$ & $\phn40.4$ \\
1b(r) & \qa &  $35174$ & $3.92$ & $0.426$ & $0.519$ & $21.6\phn$ & $\phn84.0$ \\
1b(r) & \qb &  $31916$ & $3.13$ & $0.333$ & $0.399$ & $31.3\phn$ & $\phn33.0$ \\
1b(r) & \qc &  $45518$ & $3.47$ & $0.321$ & $0.336$ & $\phn8.97$ & $\phn23.9$ \\
1b(r) & \qd &  $15710$ & $4.06$ & $0.384$ & $0.403$ & $73.8\phn$ & $137\phd\phn$ \\
\\
2b(r) & All & $135995$ & $17.9$ & $0.495$ & $0.163$ & $\phn2.47$ & $\phn4.65$ \\
2b(r) & \qa &  $35174$ & $18.4$ & $0.460$ & $0.132$ & $10.4\phn$ & $\phn9.17$ \\
2b(r) & \qb &  $31916$ & $18.4$ & $0.476$ & $0.143$ & $\phn7.61$ & $\phn4.15$ \\
2b(r) & \qc &  $45518$ & $17.8$ & $0.534$ & $0.202$ & $\phn9.71$ & $23.8\phn$ \\
2b(r) & \qd &  $15710$ & $16.7$ & $0.468$ & $0.152$ & $11.0\phn$ & $16.9\phn$ \\

\hline
\multicolumn{8}{c}{$\theta = 2.00^\circ$} \\
\hline
1a(r) & All & $131804$ & $236$ & $0.823$ & $0.131$ & $\phn2.51$ & $\phn6.24$ \\
1a(r) & \qa &  $31218$ & $258$ & $0.852$ & $0.173$ & $\phn8.49$ & $13.7\phn$ \\
1a(r) & \qb &  $29022$ & $217$ & $0.805$ & $0.117$ & $12.6\phn$ & $14.0\phn$ \\
1a(r) & \qc &  $44054$ & $234$ & $0.796$ & $0.099$ & $15.2\phn$ & $19.7\phn$ \\
1a(r) & \qd &  $13105$ & $258$ & $0.803$ & $0.096$ & $17.9\phn$ & $18.5\phn$ \\
\\
1b(r) & All & $131804$ & $14.0$ & $0.525$ & $0.245$ & $\phn4.93$ & $20.2$ \\
1b(r) & \qa &  $31218$ & $15.5$ & $0.590$ & $0.318$ & $31.2\phn$ & $69.8$ \\
1b(r) & \qb &  $29022$ & $12.3$ & $0.453$ & $0.192$ & $86.6\phn$ & $73.0$ \\
1b(r) & \qc &  $44054$ & $13.8$ & $0.458$ & $0.174$ & $30.7\phn$ & $37.3$ \\
1b(r) & \qd &  $13105$ & $16.5$ & $0.531$ & $0.215$ & $45.1\phn$ & $49.8$ \\
\\
2b(r) & All & $131804$ & $71.6$ & $0.626$ & $0.086$ & $\phn8.77$ & $\phn7.63$ \\
2b(r) & \qa &  $31218$ & $73.5$ & $0.578$ & $0.063$ & $20.8\phn$ & $16.8\phn$ \\
2b(r) & \qb &  $29022$ & $74.1$ & $0.593$ & $0.068$ & $16.0\phn$ & $14.0\phn$ \\
2b(r) & \qc &  $44054$ & $71.5$ & $0.678$ & $0.121$ & $17.0\phn$ & $24.2\phn$ \\
2b(r) & \qd &  $13105$ & $65.5$ & $0.573$ & $0.068$ & $24.6\phn$ & $23.6\phn$ \\

\hline
\multicolumn{8}{c}{$\theta = 4.00^\circ$} \\
\hline
1a(r) & All & $131467$ & $\phn945$ & $0.879$ & $0.071$ & $\phn3.62$ & $\phn4.43$ \\
1a(r) & \qa &  $24923$ & $1083$ & $0.906$ & $0.103$ & $11.3\phn$ & $11.0\phn$ \\
1a(r) & \qb &  $24505$ & $\phn888$ & $0.863$ & $0.059$ & $\phn8.63$ & $\phn8.88$ \\
1a(r) & \qc &  $42890$ & $\phn929$ & $0.843$ & $0.043$ & $18.9\phn$ & $19.7\phn$ \\
1a(r) & \qd &   $9743$ & $1024$ & $0.806$ & $0.025$ & $25.8\phn$ & $26.8\phn$ \\
\\
1b(r) & All & $131467$ & $56.2$ & $0.648$ & $0.126$ & $\phn27.1$ & $25.4$ \\
1b(r) & \qa &  $24923$ & $66.2$ & $0.726$ & $0.186$ & $\phn62.8$ & $64.0$ \\
1b(r) & \qb &  $24505$ & $50.7$ & $0.548$ & $0.077$ & $106\phd\phn$ & $91.7$ \\
1b(r) & \qc &  $42890$ & $55.2$ & $0.545$ & $0.069$ & $\phn49.1$ & $43.5$ \\
1b(r) & \qd &   $9743$ & $65.6$ & $0.516$ & $0.050$ & $\phn90.6$ & $96.0$ \\
\\
2b(r) & All & $131467$ & $287$ & $0.716$ & $0.0396$ & $11.2$ & $10.6$ \\
2b(r) & \qa &  $24923$ & $296$ & $0.659$ & $0.0257$ & $39.1$ & $35.4$ \\
2b(r) & \qb &  $24505$ & $297$ & $0.650$ & $0.0242$ & $38.6$ & $42.5$ \\
2b(r) & \qc &  $42890$ & $287$ & $0.780$ & $0.0682$ & $19.6$ & $22.5$ \\
2b(r) & \qd &   $9743$ & $249$ & $0.611$ & $0.0225$ & $35.3$ & $37.5$ \\

\enddata
\end{deluxetable*}

\begin{deluxetable*}{lc rccc cc}
\tablewidth{0pt}
\tablecaption{\label{tres-3dfvn}
$r$-band 3D Counts-in-cells $f_{V}(N)$
}
\tablehead{
\colhead{Sample} & \colhead{Quadrant} & \colhead{Cells} & \colhead{$\overline{N}$} & \colhead{$b$} & \colhead{$g$} & \colhead{$x\times10^{-5}$} & \colhead{$x\times10^{-5}$} \\
 & & & & & & \colhead{(GQED)} & \colhead{(NBD)} 
}
\startdata
\hline
\multicolumn{8}{c}{$r = 6.0 h^{-1}$ Mpc} \\
\hline
1a(r) & All & $141612$ & $4.10$ & $0.661$ & $1.88$ & $338$ & $\phn54.9\phn$ \\
1a(r) & \qa &  $35560$ & $4.46$ & $0.679$ & $1.95$ & $295$ & $115\phd\phn\phn$ \\
1a(r) & \qb &  $32542$ & $3.70$ & $0.650$ & $1.94$ & $343$ & $\phn38.8\phn$ \\
1a(r) & \qc &  $47263$ & $4.06$ & $0.644$ & $1.70$ & $433$ & $\phn\phn7.94$ \\
1a(r) & \qd &  $15545$ & $4.53$ & $0.686$ & $2.02$ & $241$ & $193\phd\phn\phn$ \\
\\
1b(r) & All & $141612$ & $0.243$ & $0.260$ & $3.40$ & $1.01\phn$ & $0.451$ \\
1b(r) & \qa &  $35560$ & $0.265$ & $0.277$ & $3.44$ & $1.19\phn$ & $1.59\phn$ \\
1b(r) & \qb &  $32542$ & $0.212$ & $0.248$ & $3.64$ & $0.270$ & $0.763$ \\
1b(r) & \qc &  $47263$ & $0.240$ & $0.232$ & $2.90$ & $3.75\phn$ & $0.374$ \\
1b(r) & \qd &  $15545$ & $0.290$ & $0.285$ & $3.30$ & $3.76\phn$ & $1.12\phn$ \\
\\
2b(r) & All & $129135$ & $0.356$ & $0.301$ & $2.94$ & $3.58$ & $0.722\phn$ \\
2b(r) & \qa &  $34172$ & $0.365$ & $0.298$ & $2.82$ & $8.34$ & $0.0680$ \\
2b(r) & \qb &  $30671$ & $0.361$ & $0.305$ & $2.97$ & $1.11$ & $4.28\phn\phn$ \\
2b(r) & \qc &  $43564$ & $0.354$ & $0.299$ & $2.93$ & $5.27$ & $0.169\phn$ \\
2b(r) & \qd &  $15448$ & $0.344$ & $0.306$ & $3.12$ & $1.89$ & $1.85\phn\phn$ \\

\hline
\multicolumn{8}{c}{$r = 12.0 h^{-1}$ Mpc} \\
\hline
1a(r) & All & $134349$ & $32.5$ & $0.790$ & $0.668$ & $30.9$ & $\phn8.72$ \\
1a(r) & \qa &  $29558$ & $35.5$ & $0.802$ & $0.693$ & $38.0$ & $11.3\phn$ \\
1a(r) & \qb &  $28014$ & $29.4$ & $0.784$ & $0.694$ & $35.1$ & $12.7\phn$ \\
1a(r) & \qc &  $44707$ & $32.3$ & $0.777$ & $0.590$ & $32.4$ & $\phn6.67$ \\
1a(r) & \qd &  $12145$ & $36.9$ & $0.817$ & $0.780$ & $23.1$ & $47.3\phn$ \\
\\
1b(r) & All & $134349$ & $1.90$ & $0.441$ & $1.16$ & $19.0$ & $30.8\phn$ \\
1b(r) & \qa &  $29558$ & $2.12$ & $0.472$ & $1.22$ & $73.5$ & $33.4\phn$ \\
1b(r) & \qb &  $28014$ & $1.64$ & $0.415$ & $1.17$ & $16.1$ & $20.7\phn$ \\
1b(r) & \qc &  $44707$ & $1.89$ & $0.406$ & $0.97$ & $22.8$ & $\phn9.39$ \\
1b(r) & \qd &  $12145$ & $2.39$ & $0.496$ & $1.23$ & $22.4$ & $68.0\phn$ \\
\\
2b(r) & All & $130665$ & $2.86$ & $0.497$ & $1.03$ & $\phn78.1$ & $\phn5.25$ \\
2b(r) & \qa &  $32557$ & $2.93$ & $0.494$ & $0.99$ & $182\phd\phn$ & $\phn8.41$ \\
2b(r) & \qb &  $29872$ & $2.92$ & $0.494$ & $0.99$ & $114\phd\phn$ & $\phn5.85$ \\
2b(r) & \qc &  $43578$ & $2.87$ & $0.501$ & $1.05$ & $\phn33.7$ & $36.2\phn$ \\
2b(r) & \qd &  $14037$ & $2.70$ & $0.486$ & $1.03$ & $\phn26.1$ & $39.1\phn$ \\

\hline
\multicolumn{8}{c}{$r = 24.0 h^{-1}$ Mpc} \\
\hline
1a(r) & All & $131132$ & $259$ & $0.860$ & $0.194$ & $\phn1.48$ & $\phn1.54$ \\
1a(r) & \qa &  $19301$ & $292$ & $0.871$ & $0.201$ & $13.4\phn$ & $\phn7.79$ \\
1a(r) & \qb &  $21734$ & $240$ & $0.856$ & $0.198$ & $\phn5.76$ & $\phn7.24$ \\
1a(r) & \qc &  $40900$ & $251$ & $0.845$ & $0.162$ & $\phn3.01$ & $\phn3.49$ \\
1a(r) & \qd &   $9179$ & $293$ & $0.881$ & $0.236$ & $16.2\phn$ & $19.6\phn$ \\
\\
1b(r) & All & $131132$ & $14.9$ & $0.588$ & $0.328$ & $\phn4.69$ & $\phn10.8$ \\
1b(r) & \qa &  $19301$ & $17.7$ & $0.632$ & $0.359$ & $35.5\phn$ & $\phn10.2$ \\
1b(r) & \qb &  $21734$ & $13.1$ & $0.551$ & $0.301$ & $55.4\phn$ & $\phn60.8$ \\
1b(r) & \qc &  $40900$ & $14.4$ & $0.547$ & $0.270$ & $23.8\phn$ & $\phn21.8$ \\
1b(r) & \qd &   $9179$ & $18.8$ & $0.612$ & $0.300$ & $99.1\phn$ & $115\phd\phn$ \\
\\
2b(r) & All & $131353$ & $23.0$ & $0.643$ & $0.298$ & $11.5\phn$ & $\phn1.05$ \\
2b(r) & \qa &  $28282$ & $23.5$ & $0.642$ & $0.289$ & $35.8\phn$ & $12.4\phn$ \\
2b(r) & \qb &  $26826$ & $23.4$ & $0.623$ & $0.258$ & $20.6\phn$ & $\phn4.82$ \\
2b(r) & \qc &  $43829$ & $23.1$ & $0.652$ & $0.314$ & $\phn4.78$ & $12.7\phn$ \\
2b(r) & \qd &  $11447$ & $21.0$ & $0.600$ & $0.250$ & $21.8\phn$ & $23.1\phn$ \\

\enddata
\end{deluxetable*}

We summarize our results in tables \ref{tres-2dfvn} and \ref{tres-3dfvn}. As expected, our results show large differences in $\overline{N}$, $b$ and $g$ between quadrants. Comparing the observed counts-in-cells distribution between the GQED and NBD, based on the least squares distance alone, $f_V(N)$ for 2D projected cells tend to follow the GQED while $f_V(N)$ for 3D spherical cells tend to follow the NBD. However, in most cases, the GQED and NBD both fall within the measured quadrant to quadrant variations, and we note that the observed $f_V(N)$ for individual quadrants within a sample may be closer to the GQED or NBD. This shows that the SDSS catalog is unable to distinguish between the GQED and NBD because the difference between the GQED and NBD is smaller than the quadrant to quadrant variations. Since the NBD has been shown to be unphysical~\citep{1996ApJ...460...16S}, on physical grounds we use the GQED for further analysis.

\subsubsection{Comparison between redshift ranges}
Comparing the values of $\overline{N}$ and $b$, between redshift ranges, we note that the values of $\overline{N}$ and correspondingly $b$ are lower in the low redshift range than in the high redshift range for the same magnitude cutoff~(samples 1b and 2b). This is despite the presence of the SDSS ``great wall'' in the low redshift range.

On a closer look, we note that the ``great wall'' spans the region between $0.065 \leq z \leq 0.09$, $140^\circ \leq \alpha \leq 210^\circ$ and $-3^\circ \leq \delta \leq 6^\circ$~\citep{2006ChJAA...6...35D} which is a small fraction of the SDSS footprint. The observed quadrant to quadrant variations of $f_V(N)$ do not seem to depart much from the GQED, so the presence of a large supercluster probably does not affect the observed form of $f_V(N)$. This result is in agreement with an earlier analysis of the 2MASS catalog by \cite{2005ApJ...626..795S} where it was found that the inclusion of the Shapley supercluster did not make a large difference on the agreement with the GQED. Therefore these large superclusters may be natural consequences of gravitational interactions among galaxies.

Since the ``great wall'' probably does not make a large contribution to the counts-in-cells statistics, the difference in $\overline{N}$ and $b$ between the high and low redshift ranges could be due to evolution or selection. Further pursuit of these possibilities requires more detailed models. Here we note that the difference is consistent across quadrants which rules out cosmic variance as a dominant cause. We also note that the difference in lookback time between the middle of the low and high redshift ranges is about 0.7 Gyr, which is close to the amount of time a merging pair of galaxies takes to merge~\citep{2009MNRAS.399L..16C} so we can expect that the effect of galaxy mergers might make a difference between the low and high redshift range.

\subsubsection{Comparison between magnitude cufoffs}
\begin{figure}
\begin{center}
\includegraphics[width=\floatwidth]{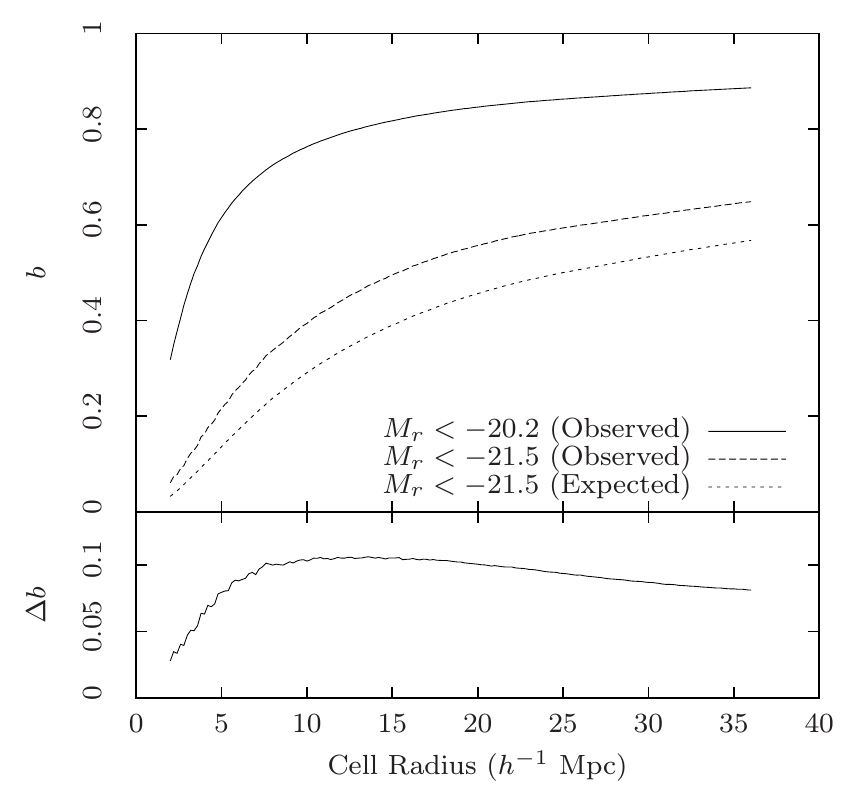}
\end{center}
\caption{\label{fres-bscomp}
The value of $b$ as a function of cell radius for bright and faint galaxies. The bottom plot is the difference between the observed and expected value of $b$ for randomly selected bright galaxies. For a given size cell, the variations between quadrants range over $2.5\%$ for the complete sample and $10\%$ for the brighter sample.
}
\end{figure}

As expected from different magnitude cutoffs, the values of $\overline{N}$ and $b$ are lower for the brighter 1b sample than the fainter 1a sample. We can compare the value of $b$ for the brighter sample with the expected value of $b$ if the brighter sample is a randomly selected sample from a fainter parent sample using~\citep{1992ApJ...396..430L}:
\begin{equation} \label{eres-bsubs}
\left(1-b_{\mathrm{sample}}\right)^2 = \frac{\left(1-b_{\mathrm{parent}}\right)^2} {1-\left(1-\frac{\overline{N}_{\mathrm{sample}}} {\overline{N}_{\mathrm{parent}}}\right)\left(2-b_{\mathrm{parent}}\right) b_{\mathrm{parent}}}.
\end{equation}
This allows us to compare the clustering of the brighter 1a sample with the fainter 1b sample. We find that for all samples, the brighter galaxies from the 1b samples have a higher value of $b$ than would be expected if they were a random subsample of the 1a sample. This means that bright galaxies are more strongly clustered than fainter galaxies which may be a result of brighter galaxies being concentrated around cluster centers, or in dark matter haloes. If so, it suggests an upper limit of about $10 h^{-1}$ Mpc for the size of these haloes. We summarize the comparisons in table \ref{tres-bscomp} and plot the comparison of $b$ for 3D cells over a range of cell sizes in figure \ref{fres-bscomp}.

\begin{deluxetable}{lcc | lcc}
\tablewidth{\floatwidth}
\tablewidth{0pt}
\tablecaption{\label{tres-bscomp}
Examples of observed and expected values of $b$ for bright randomly selected subsamples
}
\tablehead{
\multicolumn{3}{c}{2D cells} & \multicolumn{3}{c}{3D cells} \\
\colhead{$\theta$ ($^\circ$)} & \colhead{$b$ (obs.)} & \colhead{$b$ (exp.)} &
\colhead{$r$ ($h^{-1}$ Mpc)} & \colhead{$b$ (obs.)} & \colhead{$b$ (exp.)}
}
\startdata
1.0 & 0.374 & 0.256 &  6.0 & 0.260 & 0.171 \\
2.0 & 0.525 & 0.406 & 12.0 & 0.441 & 0.336 \\
4.0 & 0.648 & 0.553 & 24.0 & 0.588 & 0.492 \\
\enddata
\end{deluxetable}

\section{Discussion and Conclusion}
In this paper we have compared the counts-in-cells distribution of galaxies $f_V(N)$ with two theoretical models and found that the observed distribution has large field to field variations which may be as great as $20\%$ across quadrants. These large variations essentially mean that there is a considerable amount of cosmic variance in the data, and that the galaxies in different quadrants are not identically distributed. This also means that errors determined using the jackknife procedure will underestimate the true range of variation in the data because different subsets of the data are not identically distributed. We see the existence of these subregions of different local density from the bumps in the counts-in-cells distribution for large cells. We note that as shown by \citet{1990ApJ...353..354C}, these bumps may be the result of regions of different local density.

As suggested by \citet{2009A&A...508...17S}, a larger survey volume will show less effects of cosmic variance, and an earlier analysis of the 2MASS catalog by \citet{2005ApJ...626..795S} provides a hint that this may be the case. In the 2MASS analysis, \citet{2005ApJ...626..795S} found less variation between quadrants, on the order of $5\%$ instead of the $20\%$ we have found in the SDSS. We note that after excluding regions close to the galactic plane using $|l| < 20^\circ$, the 2MASS catalog covers two-thirds of the sky, a coverage that is more than three times that of the SDSS.

Although the SDSS ``great wall'' is contained within the low redshift range, it covers only a small fraction of the SDSS footprint in a stripe within $6^\circ$ of the celestial equator. For this reason we do not see much difference between the low redshift range and the high redshift range beyond differences in $\overline{N}$ because the overdensity in the ``great wall'' is not large enough to dominate over the rest of the survey. Indeed, because the ``great wall'' is most likely porous in three dimensions, it is not clear that ``great wall'' is an accurate description.

Our comparison of $\gamma$ for 2D projected cells and 3D spherical cells in redshift space shows that there is a difference in the exponent $\gamma$ of the two-point correlation function between the 2D and 3D sample. The 3D samples have a value of $\gamma$ that is lower than that for the 2D samples because the 3D samples in redshift space are affected by peculiar velocity distortions which change the apparent clustering in redshift space. Using the well-known relation between the projected correlation function and the real space correlation function we find that the difference in $\gamma$ between the 2D and 3D samples is a measure of the redshift space distortions. Our findings for the value of $\gamma$ suggest that the difference between the projected sample and redshift space sample is about $0.2 \sim 0.3$ in agreement with work by \citet{1994ApJ...425....1F} and \citet{2003MNRAS.346...78H} using earlier catalogs.

Comparing the low redshift and high redshift range, we find that there is a large difference in $\overline{N}$ between the low and high redshift range for samples with the same magnitude cutoff which may be caused by galaxy mergers. We also find that brighter galaxies are more strongly clustered than a random subset of all galaxies in the low redshift range, which may be a result of bright galaxies clustering around cluster centers or dark matter haloes.

The analysis of the counts-in-cells distribution shows that the observed $f_V(N)$ may follow the GQED or NBD, and 2D projected cells generally prefer the GQED while 3D spherical cells often prefer the NBD. However, both distributions are generally within the range of quadrant to quadrant variations so we can conclude that both the GQED or NBD are in agreement with observations.

Since the GQED is physically motivated while the NBD was found to be unphysical, we can reject the NBD as a physically complete description of galaxy clustering. This conclusion is not at odds with the results of \citet{2005ApJ...635..990C} because although the error bars in \citet{2005ApJ...635..990C} excludes the GQED, jackknife errors were used which, as our results suggest, underestimate the true range of variability in the data.

We conclude that while the SDSS is an excellent sample, there remains a considerable amount of cosmic variance that will probably require an all-sky survey to resolve. Nevertheless, the counts-in-cells analysis of the SDSS data has shown that observations of $f_V(N)$ agree with the GQED within cosmic variance.

\acknowledgements

We thank Matt Malkan and Gerry Gilmore for very helpful discussions on this subject.

Funding for the creation and distribution of the SDSS Archive is provided by the Alfred P. Sloan Foundation, the Participating Institutions, the National Aeronautics and Space Administration, the National Science Foundation, the US Department of Energy, the Japanese Monbukagakusho, and the Max Planck Society. The SDSS Web site is at \url{http://www.sdss.org}.

The SDSS is managed by the Astrophysical Research Consortium (ARC) for the Participating Institutions. The Participating Institutions are the University of Chicago, Fermilab, the Institute for Advanced Study, the Japan Participation Group, Johns Hopkins University, the Korean Scientist Group, Los Alamos National Laboratory, the Max-Planck-Institute for Astronomy (MPIA), the Max-Planck-Institute for Astrophysics (MPA), New Mexico State University, University of Pittsburgh, Princeton University, the US Naval Observatory, and the University of Washington.

\end{document}